\documentclass[bibyear]{aa}  
\usepackage{graphicx}
\usepackage{color}

\usepackage[colorlinks=true, linkcolor = blue, urlcolor = magenta, citecolor = blue]{hyperref}

\usepackage{txfonts}

\usepackage{tabularx}

\usepackage{newtxtext}
\usepackage[varvw]{newtxmath}
\usepackage{soul}

\usepackage{orcidlink}

\usepackage{natbib}
\bibpunct{(}{)}{;}{a}{}{,} 
\usepackage{bm}
\usepackage[normalem]{ulem}

\usepackage{placeins}
\usepackage{float}
\usepackage{multicol}

\begin{document} 

   \title{Testing the thermal Sunyaev-Zel'dovich power spectrum of a halo model using hydrodynamical simulations}
   
   \author{Emma Ayçoberry
          \inst{1}\orcidlink{0000-0002-9235-1195}
          \and
          Pranjal R. S.,
          \inst{2}\orcidlink{0000-0003-3714-2574}
          \and
          Karim Benabed
          \inst{1}
          \and
          Yohan Dubois
          \inst{1}\orcidlink{0000-0003-0225-6387}
          \and 
          Elisabeth Krause
          \inst{2,3}\orcidlink{0000-0001-8356-2014} 
          \and \\
          Tim Eifler
          \inst{2,3}\orcidlink{0000-0002-1894-3301}
          }

   \institute{
            Institut d'Astrophysique de Paris, UMR 7095, CNRS \& Sorbonne Universit\'e, 98 bis boulevard Arago, F-75014 Paris, France
        \and
            Department of Astronomy and Steward Observatory, University of Arizona, 933 North Cherry Avenue, Tucson, AZ 85721, USA
        \and
            Department of Physics, University of Arizona, 1118 E. Fourth Street, Tucson, AZ 85721, USA
            \medbreak
        \email{emma.aycoberry@iap.fr}
   }

   \date{Received XXX; accepted YYY}

  \abstract
   {Statistical properties of large-scale cosmological structures serve as powerful tools for constraining the cosmological properties of our Universe.
   Tracing the gas pressure, the thermal Sunyaev-Zel'dovich (tSZ) effect is a biased probe of mass distribution and, hence, can be used to test the physics of feedback or cosmological models.
   Therefore, it is crucial to develop robust modelling of hot gas pressure for applications to tSZ surveys.
   Since gas collapses into bound structures, it is expected that most of the tSZ signal is within halos produced by cosmic accretion shocks.
   Hence, simple empirical halo models can be used to predict the tSZ power spectra.
   In this study, we employed the \texttt{HMx} halo model to compare the tSZ power spectra with those of several hydrodynamical simulations: the \texttt{Horizon} suite and the \texttt{Magneticum} simulation.
   We examine various contributions to the tSZ power spectrum across different redshifts, including the one- and two-halo term decomposition, the amount of bound gas, the importance of different masses, and the electron pressure profiles.
   Our comparison of the tSZ power spectrum reveals discrepancies between the halo model and cosmological simulations that increase with redshift.
   We find a 20\% to 50\% difference between the measured and predicted tSZ angular power spectrum over the multipole range $\ell = 10^3 - 10^4$.
   Our analysis reveals that these differences are driven by the excess of power in the predicted two-halo term at low $k$ and in the one-halo term at high $k$.
   At higher redshifts ($z \sim 3$), simulations indicate that more power comes from outside the virial radius than from inside, suggesting a limitation in the applicability of the halo model.
   We also observe differences in the pressure profiles, despite the fair level of agreement on the tSZ power spectrum at low redshift with the default calibration of the halo model.
   In conclusion, our study suggests that the properties of the halo model need to be carefully controlled against real or mock data to be proven useful for cosmological purposes.}
    

    \titlerunning{A comparison of the tSZ properties in the \texttt{Horizon} and \texttt{Magneticum} suite of hydrodynamical simulations}
    \authorrunning{Emma Ayçoberry et al.}

   \maketitle

\section{Introduction}

The thermal Sunyaev-Zel'dovich (tSZ) effect \citep{Sunyaev1970} arises when cosmic microwave background (CMB) photons are inverse Compton scattered by hot electrons, which leads to an energy shift of the CMB photons. The tSZ effect thus manifests as a distortion of the CMB black-body spectrum. By measuring this distortion, it is possible to infer astrophysical properties of the hot gas in galaxy clusters, and, in particular, their pressure on which the tSZ signals scale, as well as cosmological information.

It is possible to observe this effect on wide fields to create tSZ maps with Planck \citep{Planck_2015} and the South Pole Telescope (SPT) \citep{Bleem_2022}, for example, or on individual galaxy clusters, as done by Planck \citep{Planck_2013_SZsource}, Atacama Cosmology Telescope (ACT) \citep{ACT_2010}, and SPT \citep{Plagge_2010} for example and will be studied with the New IRAM Kids Arrays (NIKA2) ground-based telescope \citep{Perotto_2023}. As tSZ data become better resolved (with reduced levels of noise and systematics), it is important to have a robust modelisation because it is a foreground for the CMB but also a probe of the distribution of the baryonic matter which can help us to obtain better astrophysical and cosmological constraints. On the cosmological side, the amplitude of the tSZ is extremely sensitive to $\sigma_8$ \citep[e.g.][]{Komatsu_1999, Refregier_2000, Seljak_2001, Komatsu_2002, McCarthy_2014, Bolliet_2018}. It can also be used to study early dark energy model \citep[e.g.][]{Sadeh_2007, Waizmann_2009}. On the other hand, the tSZ is sensitive to astrophysics phenomena such as active galactic nuclei (AGN) feedback that redistributes mass and modifies the pressure of the hot plasma in massive halos \citep[e.g.][]{McCarthy_2014,McCarthy_2023,Le-Brun_2015,Spacek_2018,Lee_2022,Moser_2022,Pandey_2023}.

To be able to extract even more information, the tSZ has been used in correlation with other probes. Many combined cross-correlation analyses have been performed, for example, the correlation between the tSZ signal and the weak lensing signal has been widely used to constrain cosmological parameters and nuisance parameters \citep[e.g.][]{Van-Waerbeke_2014,Ma_2015,Osato_2020,Troster_2022}. More recently, \cite{Fang_2023} developed a joint halo model to correlate tSZ, weak lensing, CMB lensing, and galaxy density (in total ten different two-point functions) to obtain even tighter constraints jointly on cosmological parameters and astrophysical parameters. Incorporating the tSZ probe is beneficial due to its sensitivity to baryonic physics. From this paper, one can note that, by including cross-correlations with additional tracers, the figure-of-merit can be improved by a factor of two and reduced notably the fraction of sky needed. By adding stronger priors on the halo model, in particular on the parameters required to model the tSZ pressure profile, one can gain even more on the figure-of-merit, which emphasizes the need for more robust tSZ models.

Originally, the tSZ power spectrum was modelled through the Press-Schechter formalism \citep[e.g.][]{Komatsu_1999, Refregier_2000,Seljak_2001}. First studies with this formalism only account for correlation within clusters (known as the one-halo term) and \cite{Komatsu_1999} added for the first time the contribution of the correlation among clusters (known as the two-halo term). To improve the prediction, the Press-Schechter formalism is replaced by a halo model that depends mostly on the distribution of halos, halo bias, and electron pressure profile. \cite{Refregier_2002} made the first comparison between the tSZ power spectrum obtained with such a model and measurement in simulation. Once the limitations of the simulation were taken into account, the prediction from the halo model was in good agreement with the simulation. Many works continue to model the tSZ power spectrum with different parametric choices in a halo model framework to improve the agreement of the model with improved simulations and measurement. For cosmological analysis, one of the most used halo model is \texttt{HMx}, developed by \cite{Mead_2020} \citep[e.g. in][]{Troster_2022}. It is also possible to study the tSZ with non-parametric modelisation with machine learning techniques as done with the \texttt{camels} simulations \citep{Moser_2022} or with the \texttt{the three hundred project} \citep{Ferragamo_2023} for example.

For this work, we focused on and used the \texttt{HMx} halo model developed by \cite{Mead_2020}. This halo model was calibrated on the \texttt{BAHAMAS} simulation (\citealp{McCarthy_2017-bahamas}, see a small description of the simulation in Sect.~\ref{sec:bahamas} of this paper) at the level of the response power spectrum, and we used the model calibrated for the matter-pressure power spectrum. We explored how the predictions and components of this model compare with the \texttt{Horizon} suite~\citep{Dubois_2014, Dubois_2016} and the \texttt{Magneticum}~\citep{Dolag_2016} simulation. This study allowed us to extract the tSZ signal of the simulations to gauge the uncertainties associated with subgrid modelling. We also explored the importance of different mass bins and redshift ranges, to understand better where the modelisation needs to be improved, in particular for the correlation between CMB lensing and tSZ signal of clusters.

This paper is organised as follows: in Sect.~\ref{sec:model} we describe the \texttt{HMx} halo model. We continue with a description of the \texttt{Horizon} suite and \texttt{Magneticum} simulation in Sect.~\ref{sec:simu}. In Sect.~\ref{sec:res} we present the results of the comparison of the power spectrum and angular power spectrum. In Sect.~\ref{sec:discussion} we discuss the different properties that can impact the difference between prediction and measurement. Finally, we draw our conclusions in Sect.~\ref{conclu}.

\section{Halo model framework}
\label{sec:model}
We recall in the following the main assumptions and properties of a classical halo model and then describe the specifics of \texttt{HMx}.

\subsection{Halo model}
A halo model assumes that all matter in the universe is partitioned into spherical and symmetrical halos of different masses and sizes. Depending on the scale of interest, the statistical properties of the matter distribution will depend on the overall distribution of halos in the universe (large scale) or on the distribution of matter within a given halo, integrated over the distribution of halo sizes and masses (small scale). Such a model relies on a limited number of components, including the properties of the matter distribution at large scales in the linear regime (given by the cosmological model), the distribution of halo mass (or halo mass function) $n(M)$, the halo bias $b(M)$, and the halo profile. These components must be modelled and can be calibrated using simulations or data. Starting from this main idea, it is possible to extend this formalism to other tracers of large-scale structures, provided that the halo properties as seen by those tracers and their correlations with the matter distribution are known.

In this context, the power spectra of the different tracers of the large-scale structure (LSS) are often separated into their one-halo (probing the intra-halo statistical properties of the tracer, typically at small scales) and two-halo parts (probing the distribution of halos, often at large scales) $P_{uv}(k) = P_{{\rm 1H},uv}(k) + P_{{\rm 2H},uv}(k)$, where $u$ and $v$ are two three-dimensional (3D) fields.  

In detail, the one- and two-halo power spectrum of a given tracer of the LSS follow
\begin{eqnarray}
    P_{{\rm 1H},uv}(k) &=& \int_0^\infty W_u(M,k)W_v(M,k)n(M){\rm d}M,
\\
    P_{{\rm 2H},uv}(k) &=& P_\text{lin}(k)\prod_{i=u}^{v}\left[ \int_0^\infty b(M)W_i(M,k)n(M){\rm d}M \right],
\end{eqnarray}
where $P_{\rm lin}$ is the linear matter power spectrum and the Fourier transform of the field $u$ is
\begin{equation}
    W_u(M,k) = \int_0^\infty 4\pi \hat r^2 \frac{\sin(k\hat r)}{k\hat r}\theta_u(M,\hat r){\rm d}\hat r,
\end{equation}
with $\theta_u(M,\hat r)$ the averaged radial profile for the field $u$, function of mass $M$ and comoving radius $\hat r$. 

While the matter, pressure, and matter-pressure 3D power spectra can be measured in simulations and are required to compute the correlation of tSZ with other LSS tracers, they are usually not directly observable. What can be measured easily in microwave sky maps are angular power spectra $C_{uv}(\ell)$ which are projected along the line of sight. From the 3D power spectra, one can compute the angular power spectrum $C_{uv}(\ell)$ 
\begin{equation} \label{eq:Cl}
    C_{uv}(\ell) = \int_0^{\hat r_{\rm H}} \frac{X_u(\hat r) X_v(\hat r)}{f_k^2(\hat r)} P_{uv}(k(\hat r),z(\hat r)){\rm d}\hat r,
\end{equation}
where $\ell$ is the multipole moment, $\hat r_{\rm H}$ the Hubble radius, $f_k(\hat r)$ the comoving angular-diameter distance (for a flat universe we thus have $f_k(\hat r) = \hat r$), $z(\hat r)$ is the redshift at comoving distance $\hat r$, and $k(\hat r) = (\ell+1/2)/f_k(\hat r)$. For the Compton-$y$ parameter, the projection kernel $X_y$ is:
\begin{equation} \label{eq:Xy}
    X_y(\hat r) = \frac{\sigma_{\rm T}}{m_{\rm e}c^2}\frac{1}{a^2(\hat r)}\, ,
\end{equation}
where $a$ is the expansion factor, $\sigma_{\rm T}$ is the Thompson scattering cross-section, $m_{\rm e}$ is the electron mass, and $c$ is the speed of light.

\subsection{\texttt{HMx} parametrisation}
\label{sec:hmx}
We recall in the following the details of one of the main approaches used to approximate the power spectrum of different tracers of large-scale structures: the \texttt{HMx} model \citep{Mead_2020}. This semi-analytical model was built upon the classical halo model and included additional degrees of freedom that were fit to a suite of hydrodynamical simulations \citep[the \texttt{BAHAMAS} simulation][]{McCarthy_2017-bahamas}. The \texttt{HMx} model predicts the (cross-)power spectra of different tracers, including tSZ, within a single framework and includes physics-inspired parameters that are common to some of the different tracers. This approach is particularly useful when analysing a large number of tracers simultaneously in the context of cosmological surveys, as we demonstrate in \cite{Fang_2023}.

Different halo models for different observables and assuming different hypotheses have been developed over the years \citep[e.g.][]{Maniyar_2021,Bolliet_2023}. Each of them proposes different modelling for the components described above and different ways to calibrate the parameters of those models. The \texttt{HMx} \citep{Mead_2020} variant of these models is one of the most widely used for cosmological data analysis (e.g. for weak lensing analysis such as in \citealp{Troster_2022}). It further features several interesting features that make it particularly suitable as a base for comparison with simulations, as we will do here. \texttt{HMx} already implements several probes of the LSS which makes it particularly useful in the context of probes cross-correlations. Its implementation is also very modular, allowing for relatively easy modifications of the different model components, as well as change in ranges of integration, a feature that will be key to our comparisons, as we will discuss below. Finally, a specificity of this model is that all the parameters of the different approximations used to describe the halo properties are fitted simultaneously against the \texttt{BAHAMAS} simulation power spectra \citep[see the end of this subsection or in][for more details]{Mead_2020}. This is a different approach from one where each component and parameter of the model are fitted separately against data or simulations (for example, fitting the parameters of the halo profile against stacked halo profiles measured in simulations). This simultaneous fitting ensures the fidelity of the power spectra prediction, but it may come at the expense of the physical meaning of the parameters and their values. This can be an issue if one wants to improve the model by obtaining stronger and physically inspired priors on those parameters from data or simulations. For all of those reasons, we decided to retain \texttt{HMx} as our reference halo model in this work.

The mass function $n(M)$ adopted in \texttt{HMx} is the one from \cite{Sheth_1999}. The halo bias function was then derived from this mass function using the peak-background split formalism \citep{Mo_1996,Sheth_2001}. In the case of tSZ, the field $u$ is the electron pressure, defined as
\begin{equation} \label{eq:profile}
    P_{\rm e}(M,r) = \frac{\rho_{\rm bnd}(M,r)}{m_{\rm p} \mu_{\rm e}}k_{\rm B} T_{\rm g}(M,r), 
\end{equation}
where $m_{\rm p}$ is the proton mass, $\mu_{\rm e}$ is the mean gas-particle mass per electron divided by the proton mass, $k_{\rm B}$ is the Boltzmann constant, and $\rho_{\rm bnd}$ is the halo bound gas density and represents the gas which is inside the proper virial radius $r_{\rm v}$, defined as
\begin{equation}
    \rho_{\rm bnd} \propto \left[ \frac{\ln(1+r/r_{\rm s})}{r/r_{\rm s}} \right]^{1/(\Gamma-1)},
\end{equation}
where $\Gamma$ is the polytropic index for the gas, and $r_{\rm s}$ is the halo scale radius parameter, specified via the concentration relation $c_{\rm M}=r_{\rm v}/r_{\rm s}$

The virial mass $M_{\rm v}$ was defined as where the total mean matter density in the halo within the (proper) virial radius $r_{\rm v}$ is $\Delta_{\rm v}$ (the virial-collapse density contrast) times the critical density~:
\begin{equation}
\label{eq:Mvir}
    M_{\rm v} = \frac{4\pi}{3}r_{\rm v}^3 \Delta_{\rm v} \rho_{\rm c}(z),
\end{equation}
where $\rho_{\rm c}(z)=3H(z)^2/(8\pi G)$ is the critical density of the Universe at redshift $z$, $G$ is the gravitational constant, $H(z)$ is the Hubble expansion factor, and where $\Delta_{\rm v}$ comes from the $\Lambda$CDM fitting function of \cite{Bryan_1998}. This was the same definition of virial mass as in \cite{Mead_2020}, but we caution that their masses were measured in dark matter-only simulation whereas we used the one of the hydrodynamical simulations directly.

The gas ejected outside the virial radius by feedback processes did not contribute to the one-halo term. The formalism assumed that the gas was associated with the initial density of the associated halo but can now be far from it. It was however taken into account by adding it back to the two-halo term only, following the treatment done by \cite{Fedeli_2014} and \cite{Debackere_2020}. For more details about the treatment of the ejected gas, we refer the reader to \cite{Mead_2020}. The model further assumed that all the gas is ionized, and for the bound gas, the \cite{Komatsu_2001} profile was used to determine the gas temperature $T_{\rm g}$:
\begin{equation}
    T_{\rm g}(M,r) = T_{\rm v}(M)\frac{\ln(1+r/r_{\rm s})}{1+r/r_{\rm s}},
\end{equation}
which assumes hydrostatic equilibrium. The virial temperature $T_{\rm v}$ was defined as:
\begin{equation}
    \frac{3}{2}k_{\rm B}T_{\rm v}(M) = \alpha \frac{GM m_{\rm p}\mu_{\rm p}}{r_{\rm v}},
\end{equation}
where $\mu_{\rm p}$ is the mean gas-particle mass divided by the proton mass, $\alpha$ encapsulates deviations from a virial relation and thus acts as a hydrostatic bias (here we have $\alpha = 0.8471$). 

Such model particularly enhances the contribution of high-mass halos to the tSZ power spectrum, while low-mass halos are deficient in gas. Moreover, the amplitude of the one- and two-halo terms are more sensitive to high-mass halos, as the Fourier transform of the pressure profile $W_{\rm p}(M,k \rightarrow 0)$ is proportional to $M^{5/3}$. In comparison, the Fourier transform of the matter field $W_{\rm m}(M,k \rightarrow 0)$ scales as $M$. This can be explained by the fact that the electron pressure follows the gas density but emanates only from the highest gas-density peaks because the temperature is higher, thus boosting the electron pressure.
  
For more details about the definition of the concentration relation, the fraction of bound gas, or other model's components, we refer the reader to \cite{Mead_2020}. The default and best-fit values of the parameters of the model, that were fitted against the \texttt{BAHAMAS} simulations \citep{McCarthy_2017-bahamas}, can be found in Table~1 and Table~2 of \cite{Mead_2020}, respectively.

In detail, the fits were performed on the \texttt{BAHAMAS} simulation 3D power spectrum response between $z=0$ and $1$ and for $k$ between $0.015$ and $7 \, h^{-1}\, \rm Mpc$ for stars, matter, matter \& electron pressure, or matter, cold dark matter (CDM), gas \& stars jointly. The fits covered four different cosmological models. In the case of tSZ, the parameters were fitted on the matter-electron pressure cross-power spectrum only and used as is for the predictions of the electron pressure auto-spectrum. The \cite{Mead_2020} paper argues that this approach provides the lowest error on the pressure auto-power spectrum as the pressure auto-power spectrum is difficult to fit because of its high Poisson noise. The relative difference between the \texttt{BAHAMAS} simulation and prediction, averaged linearly over $z$ between $0$ and $1$ and logarithmically over $k$ between $0.015$ and $7\,h \,\rm Mpc^{-1}$, is of $2\,\%$ for the matter auto-power spectrum, $15\,\%$ for the matter-pressure, $25\,\%$ for the pressure auto-power spectrum. As can be seen, the prediction has a relatively low fidelity for the pressure auto-spectrum. Several reasons can explain this fact. The halo model makes several strong hypotheses, such as halos trace the underlying linear matter distribution with a linear halo bias, halo profiles are perfectly spherical with no substructure and no scatter at fixed mass, and there is nothing to prevent halos from overlapping. While these hypotheses provide a reasonable approximation of the physics at play in the case of the matter distribution, explaining the good precision in the case of the matter power spectrum and fair in the case of the cross-spectrum, they are not necessarily correct for the electron pressure distribution. Furthermore, the tSZ sensitivity to high-mass clusters also limits the predictive power for different reasons. First, because of the limited size of the simulations, calibration for high-mass clusters can be imperfect. Second, the mass scaling parametrisation must extrapolate from halo masses probed by the matter-pressure power spectrum to those probed by the pressure auto-power spectrum, which involves highest masses.

The relatively low fidelity of the model for the electron pressure auto-spectrum, the fact that the parameters lose some of their physical meaning since they were fitted to reproduce the simulation power spectra and the risk of over-fitting the model to a particular simulation and its limitations are all excellent arguments to revisit the \texttt{HMx} predictions and compare them with other simulation suites. In this work, we used the \texttt{Horizon} and \texttt{Magneticum} simulations. Similar to the approach taken by \cite{Mead_2020}, we also focused on the matter-pressure cross-spectra, which is the best predicted observable of \texttt{HMx} model we are using.

\subsection{Angular power spectrum prediction}
\label{section:angulappshalomodel}
\begin{figure*}
    \centering
    \includegraphics[width=0.9\textwidth]{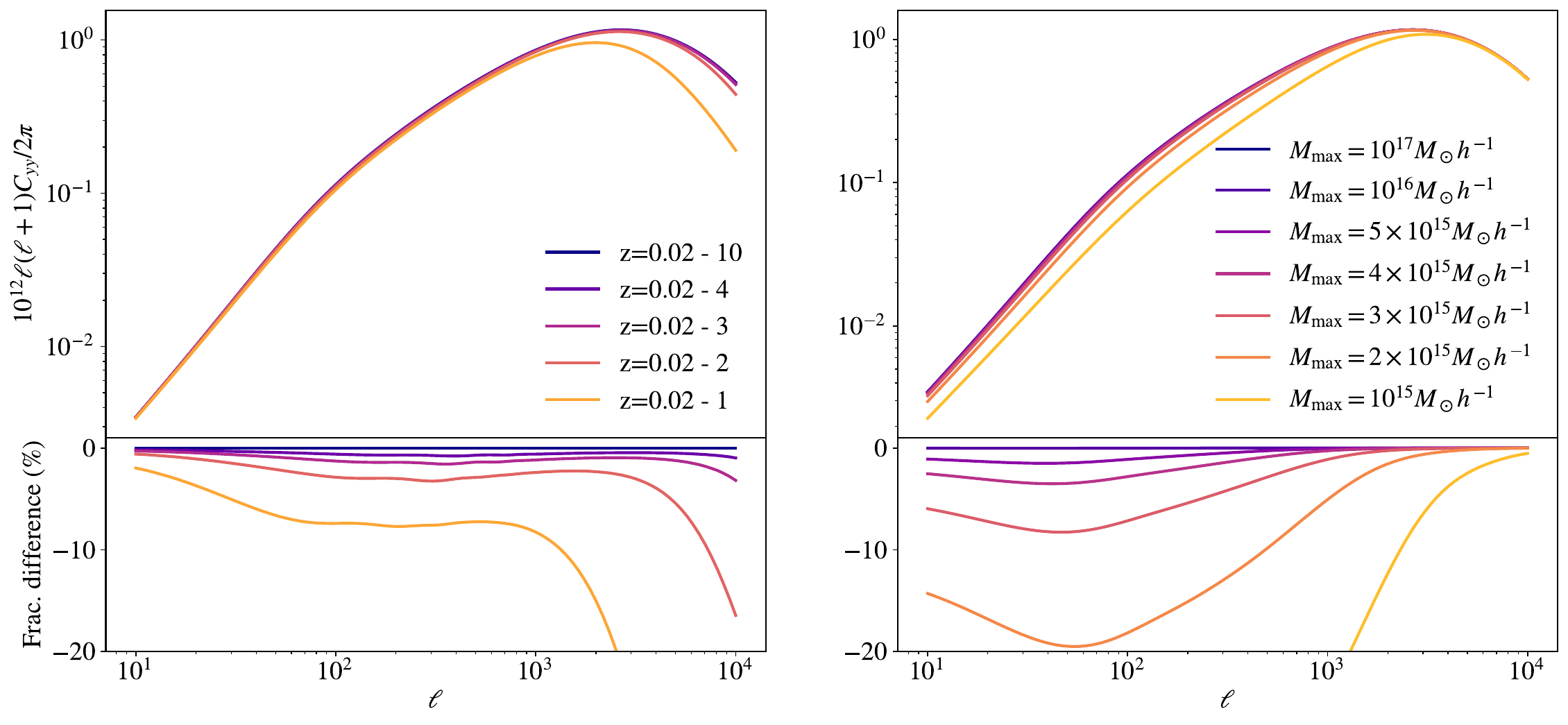}
    \caption{Predicted pressure angular power spectrum when integrating up to different redshifts in different colours (on the left) or between $z=0.02$ and $z=10$ for different maximal mass in different colours (on the right) as a function of the angular scale $\ell$.}
    \label{fig:hmcode_limber_z}
\end{figure*}

In the rest of this article, we compared the model predictions with measurements in the simulations. However, it is not useful to discuss differences occurring at redshifts that do not contribute significantly to the angular power spectra. For example, \cite{Komatsu_2002} showed that the contribution of clusters at $z>10$ is negligible to the pressure angular power spectrum. We used the predictions from \texttt{HMx} to address this question. The pressure angular power spectrum obtained using different redshift ranges of integration is shown on the left panel of Fig.~\ref{fig:hmcode_limber_z}, where we look at the angular power spectrum when integrating up to different redshifts. We can note that integration up to $z=3$ or $z=4$ captures more than $97\,\%$ of the power for $\ell$ between $10$ and $10^4$. Limiting ourselves to $z=2$ will only lead to a dramatic loss of $\sim 17\%$ of validity after $\ell=4 \times 10^3$. We have also looked at the contribution coming from the one-halo term when integrating up to a given $z$ for different $\ell$. Except when $z$ is sufficiently small and $\ell$ large, corresponding to the interior of halos, more than $90\,\%$ of the angular power spectrum comes from the one-halo term. This behaviour can be due to the change of slope in the pressure profiles of halos, or to the fact that the contrast of pressure is smaller on the inner part of halos. These two conclusions imply that the modelisation of the electron pressure profiles up to $z=4$ will impact our results and need to be well understood.

In the previous section, we noted that the power spectra are dominated by high-mass clusters. However, depending on the simulation characteristics, the number of such high-mass objects in our simulations will be limited, which we need to take into account when comparing the model predictions with the simulations. As a first step in investigating the impact of high-mass objects, we show in the right panel of  Fig.~\ref{fig:hmcode_limber_z} the contribution of the different halo masses to the power spectra, as we vary the maximum mass considered in \texttt{HMx}~\citep[see also, e.g.][]{Refregier_2000,Battaglia_2012}. Our reference here is the default maximum mass used in \texttt{HMx}: $M_\text{max} = 10^{17} \, h^{-1} \, M_\odot$. We can note that the different $\ell$ are not affected in the same way; this choice of maximal mass impacts the most the values of $\ell = 50 - 60$. We can see that integrating up to $M_\text{max} = 10^{16} \, h^{-1} \, M_\odot$ makes almost no differences. Using Gumbel statistics, \cite{Davis_2011} found that it is very unlikely to have dark matter halos with $M > 10^{16} \, h^{-1} \, M_\odot$ within a volume of $1 \, h^{-1} \, \rm Gpc$, we can extrapolate that it is also the case within the observable Universe. The prediction still changes a bit because the halo mass function predicts that it is possible to have such halos, but for analysis with real data, we will probably never use this maximal mass \citep{Holz_2012}. If we are integrating up to $M_\text{max} = 2\times10^{15} \, h^{-1} \, M_\odot$ instead of $M_\text{max} = 4\times10^{15} \, h^{-1} \, M_\odot$ we loose maximum $20\,\%$ of the signal instead of a few percent. This implies that the masses of a few $10^{15} h^{-1} M_\odot$ must be well-modelled in our prescription. 

\section{Simulations}
\label{sec:simu}
To pursue this analysis, we are interested in the comparison of the \texttt{HMx} prediction with measurement in different simulations. We thus used four simulations: \texttt{Horizon-AGN}, \texttt{Horizon-noAGN}, \texttt{Horizon-Large}, and \texttt{Magneticum} and we recall their main characteristics in this section. This allowed us to compare prediction and measurement with simulations containing different physics and implementation schemes but also to analyse for the first time the tSZ signal in the \texttt{Horizon} simulations. We also recall the main characteristics of the \texttt{BAHAMAS} simulation because they are the ones used to calibrate \texttt{HMx}, but we are not analysing them within this paper.

\subsection{\texttt{Horizon-AGN}}
The \texttt{Horizon-AGN} simulation \citep{Dubois_2014} is a cosmological hydrodynamical simulation of $100 \,h^{-1}\,\rm Mpc$ comoving volume, with $1024^3$ dark matter particles, leading to a resolution of $M_\text{DM,res} = 8.3 \times 10^7 \, M_\odot$. The simulation uses the adaptive mesh refinement code \texttt{RAMSES} \citep{Teyssier_2002} that can refine up to a minimum cell size of $\Delta x\simeq 1\,\rm kpc$ (comoving). The cosmology is a standard $\Lambda$CDM cosmology compatible with WMAP-7 \citep{Komatsu_2011} with \{$\Omega_{\rm m}, \Omega_\Lambda, \sigma_8, \Omega_{\rm b}, n_{\rm s}$\} = \{$0.272, 0.728, 0.81, 0.045, 0.967$\} and $H_0 = 70.4 \, {\rm km \, s^{-1} \, Mpc^{-1}}$. Multiple redshifts between $z=0.018$ and $z=38.3$ are available, allowing us to perform redshift space analysis. In the following, we focus our study on redshifts between $0$ and $5$. More details about the physics and refinement scheme are available in \cite{Dubois_2014}, but we summarise here the main aspects.
The simulation includes gas cooling~\citep{Sutherland93}, and a uniform UV background~\citep{Haardt96} with redshift of reionisation $z_{\rm r}=10$.
It follows star formation using a Schmidt law with $2\,\%$ star formation efficiency and the associated feedback from stellar winds, type II and type Ia supernovae~\citep{Dubois08}, as well as feedback from active galactic nuclei (AGN) powered by Bondi-Hoyle-Lyttleton accretion limited at Eddington with jet/radio or heating/quasar mode depending on the accretion rate relative to Eddington~\citep{Dubois12}.

\subsection{\texttt{Horizon-noAGN}}
The \texttt{Horizon-noAGN} simulation \citep{Dubois_2016} has the same initial conditions, sub-grid modelling, and cosmology as the \texttt{Horizon-AGN} simulation, only the physics is different. This simulation contains no black hole formation and therefore no AGN feedback.
This leads to a significant overshoot of the baryonic mass content in galaxy groups and clusters, and in particular of their gas fraction, at the high-mass end~\citep{Beckmann17,Chisari18}.

\subsection{\texttt{Magneticum}}
The \texttt{Magneticum} suite of simulations \citep{Dolag_2016} are cosmological hydrodynamical simulations with different box sizes and cosmologies. In this work we use the medium resolution \textit{Box1a} simulation of $896\, h^{-1}\,\rm Mpc$ comoving volume, with $1512^3$ dark matter and (initial) gas particle masses of $1.3\times 10^{10}\, h^{-1}\, M_\odot$ and $2.6\times 10^{9}\, h^{-1}\,M_\odot$, respectively. For analysing the properties of lower mass halos (see section~\ref{sec:pressure_profile}), we use \textit{Box2} which has a smaller volume of $352 \, h^{-1}\,\rm Mpc$ comoving volume but a better mass resolution of $6.9\times10^{8}\, h^{-1}\,M_\odot$ and $1.4\times 10^{8}\, h^{-1}\,M_\odot$ for dark matter and gas particles respectively. The simulations use the smooth particle hydrodynamics (SPH) code \texttt{P-GADGET3} \citep{Springel2005}. The boxes that we are using also follow a WMAP-7 cosmology \citep{Komatsu_2011} with $\{\Omega_{\rm m}, \Omega_{\rm b}, \sigma_8, h, n_{\rm s}\}=\{0.272, 0.0456, 0.809, 0.704, 0.963\}$. We also have access to the redshifts between $0$ and $5$. We summarise here the main physical aspects. The simulation includes gas cooling, star formation, and winds \citep{Springel_2003}. The metals, stellar population, and chemical enrichment, SN-Ia, SN-II, AGB follow \cite{Tornatore_2003, Tornatore_2006} with the new cooling tables of \cite{Wiersma_2009b}. There are also black holes and AGN feedback \citep{Hirschmann_2014}. 

\subsection{\texttt{Horizon-Large}}
We ran the \texttt{Horizon-Large} simulation specially for this work to improve the comparison between the \texttt{Horizon} and \texttt{Magneticum} suite of simulations. This particular simulation is a cosmological hydrodynamical simulation of $896\,h^{-1}\,\rm Mpc$ comoving volume (a similar box size than that of the \texttt{Magneticum} simulation described in the previous subsection), with $1024^3$ dark matter particles, leading to a resolution of $M_\text{DM,res} = 6 \times 10^{10}\, M_\odot$. The simulation uses the \texttt{RAMSES} \citep{Teyssier_2002} code and the grid is allowed to refine up to a spatial resolution of $10\,\rm kpc$ (comoving). The cosmology is the same as that of the \texttt{Horizon-AGN} and \texttt{Horizon-noAGN} simulations. The physics is simpler than the other simulations: it only contains gas cooling and UV background heating below $z_{\rm r}$ and no galactic physics (no star formation nor feedback) for computational reasons. 
It is a reasonable approximation to large-scale boxes with more complex subgrid physics (such as, e.g. \texttt{Magneticum}) as the most massive clusters that are captured in such big volumes -- and which dominate the tSZ signal (see section~\ref{sec:hmx}) -- are the most insensitive to mass redistribution due to feedback~\citep[e.g.][]{Gonzalez_2013,Le-Brun_2014, McCarthy_2018, Chisari18}. 
This simulation is also used to probe the cosmic variance with a similar numerical treatment of the hydrodynamics than in the other \texttt{Horizon} simulations. 

\subsection{BAHAMAS} \label{sec:bahamas}
The \texttt{BAHAMAS} simulations are a suite of hydrodynamical simulations of $400 \, h^{-1} \, \rm Mpc$ with the WMAP 9-yr \citep{Hinshaw_2013} and Planck 2013 \citep{Planck_2013-cosmo} cosmology. The simulations contain $2 \times 1024^3$ particles leading to a resolution of $M_\text{baryon,res} = 3.85 \times 10^9 \, h^{-1}\, M_\odot$ ($M_\text{baryon,res} = 4.45 \times 10^9 \, h^{-1} \, M_\odot$)  and $M_\text{DM,res} = 7.66 \times 10^8 \, h^{-1} \, M_\odot$ ($M_\text{DM,res} = 8.12 \times 10^8 \, h^{-1} \, M_\odot$), respectively for a WMAP-9 (Planck) cosmology. The hydrodynamic code and subgrid physics are the same as the ones in the \texttt{OWLS} \citep{Schaye_2010} and \texttt{cosmo-OWLS} \citep{Le-Brun_2014, McCarthy_2014} projects. The simulations include radiative cooling and heating, with a reionization that occurs at $z_r=9$, a star formation rate, and stellar evolution and chemical enrichment. It also contains a black hole and AGN feedback with three strengths of AGN feedback, from the smaller to the bigger: $10^{7.6}\,\rm K$, $10^{7.8}\,\rm K$, and $10^{8.0}\,\rm K$. More details about the physics are available in \cite{Schaye_2010}. 

\subsection{Mass cut to compare different simulations}
\begin{figure*}
    \centering
    \includegraphics[width=0.9\textwidth]{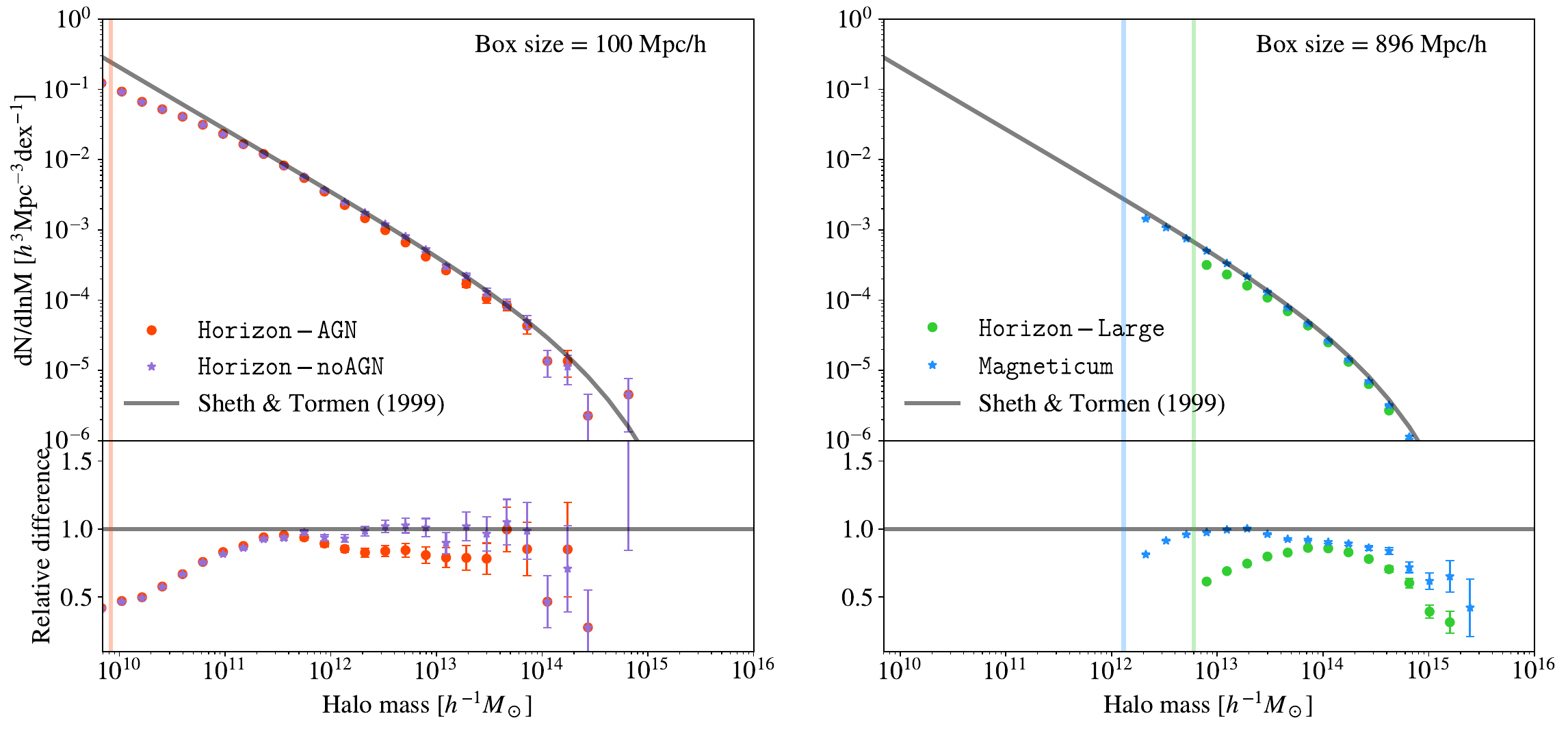}
    \caption{Halo mass function at $z=0$ for the \texttt{Horizon-AGN}, \texttt{Horizon-noAGN}, \texttt{Horizon-Large}, and \texttt{Magneticum} simulations in red, purple, green, and blue, respectively, with their associated Poisson noise as error bar. Each mass function is computed up to the maximal halo mass in the corresponding simulation. The analytical mass function from \cite{Sheth_1999} is shown in black for comparison. Vertical lines indicate the minimum halo mass accessible in each simulation, defined as $100$ times the dark matter mass resolution. The lower limits for \texttt{Horizon-AGN} and \texttt{Horizon-noAGN} are identical.}
    \label{fig:hmf}
\end{figure*}

Because the simulations have different box sizes and physics, the halo mass function can differ and, as seen on the right of Fig.~\ref{fig:hmcode_limber_z} and related text, the maximum mass chosen for the \texttt{HMx} prediction will impact our results. In Fig.~\ref{fig:hmf}, we present the halo mass function of the four simulations we are working with (\texttt{Horizon-AGN}, \texttt{Horizon-noAGN}, \texttt{Horizon-Large}, and \texttt{Magneticum}) compared with the theoretical halo mass function from \cite{Sheth_1999}, which is the one used in \texttt{HMx}. This analytical mass function has been fitted on dark matter halos. For the \texttt{Horizon} simulations, the halos were identified with the \texttt{adaptaHOP} halo finder \citep{Aubert_2004}, and we only kept the halos (and subhalos) with at least 100 particles. For \texttt{Magneticum}, the halos were identified with a standard Friends-of-Friends algorithm and subhalos with the \texttt{subfind} module \citep{Springel_2001, Dolag_2009}. We see a relatively good agreement between the different simulations, nevertheless, they all have different maximal and minimal masses because of the difference in volume and mass resolution. We observe a lack of high-mass halos (except for the last bin of \texttt{Horizon-AGN} and \texttt{Horizon-noAGN}, which is probably just a binning effect), more important in the bigger simulations, and of low-mass halos in all the simulations. The trend observed at $z=0$ is similar to the one at higher redshift. Since we are comparing an analytical function derived from dark matter-only halos with results from hydrodynamical simulations, discrepancies may be caused by the influence of baryons (cosmological accretion shocks, background UV heating, and feedback) on mass redistribution. The impact varies across different halo masses with fractional variations in good agreement with other studies \citep[see, e.g.][]{ Vogelsberger_2014,Sorini_2024}.

We recall that we used the virial mass as the definition of our halo mass (defined in Eq.~\ref{eq:Mvir}). As mentioned in Sect.~\ref{sec:model}, the maximal mass chosen in \texttt{HMx} will significantly change the prediction on the pressure auto-power spectrum, it is thus important to take them into account when comparing the results from \texttt{HMx} and the one measured in the simulations. In the following of the paper, we used $M_\text{max} = 6.4 \times 10^{14}\, h^{-1} \,M_\odot$ for \texttt{Horizon-AGN} and \texttt{Horizon-noAGN} simulations and $M_\text{max} = 2.6 \times 10^{15}\, h^{-1}\, M_\odot$ for the \texttt{Horizon-Large} and \texttt{Magneticum} simulations. These two maximal masses are the maximum mass of \texttt{Horizon-AGN} and \texttt{Magneticum}, respectively. We have checked that the prediction using the maximal mass of \texttt{Horizon-noAGN} (or \texttt{Horizon-Large}) makes almost no difference, this is why we can compare the results of \texttt{Horizon-noAGN} with the prediction obtained for \texttt{Horizon-AGN}, and the results of \texttt{Horizon-Large} with the prediction obtained for \texttt{Magneticum}.

\section{Results} \label{sec:res}
In this section, we present our results on the comparison of the predicted (angular) power spectrum from \texttt{HMx} and the one measured in the different simulations. As some differences between the predictions and the measurements are expected (see Sect.~\ref{sec:model}), we now focus on characterizing the ones that matter the most for observations.

\subsection{Power spectrum}
The first straight-forward comparison concerns the pressure and matter-pressure power spectrum. We restricted our analysis to redshifts between $0$ and $\sim 4$, as higher redshifts do not significantly contribute to the angular power spectrum, as discussed in Sect.~\ref{section:angulappshalomodel}. We focused on quantifying differences of the tSZ signal from simulations that include different physical models than \texttt{BAHAMAS} (which \texttt{HMx} is calibrated on) and understanding how they propagate to higher redshifts.

To obtain the different power spectra from the simulations, we followed the procedure described in Appendix~\ref{app:power_spectrum}. As a first check, we compared the matter auto-power spectrum of all the simulations at all redshift with the one predicted by \texttt{HMx} and we find a good agreement. The analysis reveals a $\sim 10\,\%$ difference between the measurements and predictions, with no apparent trend in $k$ or redshift. This test allows us to confirm our pipeline before moving to the analysis of pressure auto- or cross-power spectra.

\subsubsection{Pressure auto-power spectrum}
\begin{figure*}
    \centering
    \includegraphics[width=0.9\textwidth]{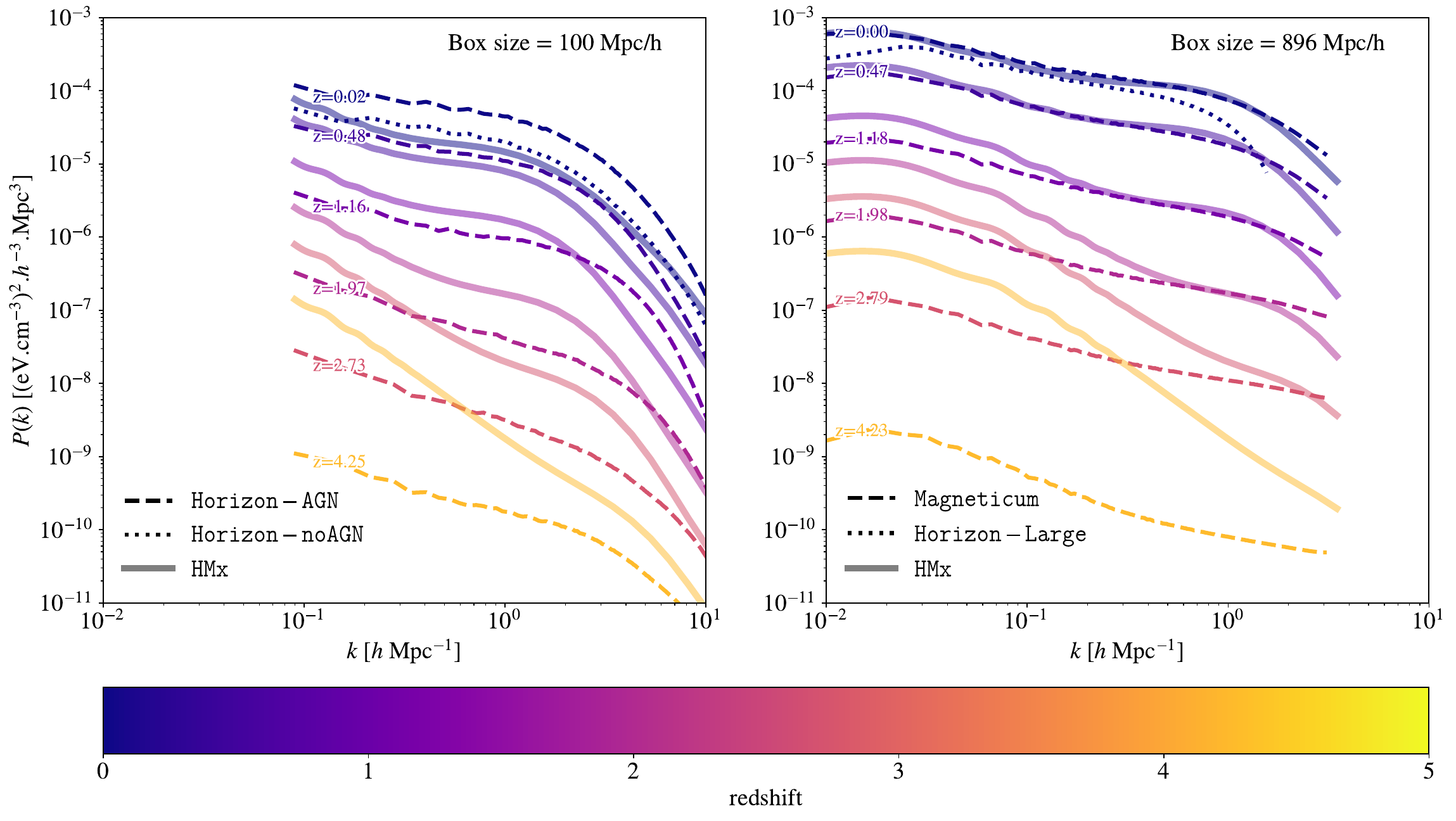}
    \caption{Pressure auto-power spectrum as a function of redshift. The left panel shows the results for simulations with a box size of $100\, h^{-1}\,\rm Mpc$, thus \texttt{Horizon-AGN} in dashed line compared to \texttt{HMx} in solid line, and \texttt{Horizon-noAGN} at $z=0$ in dotted line. The right panel shows the results for simulations with a box size of $896\, h^{-1}\,\rm Mpc$, thus \texttt{Magneticum} in dashed line compared to \texttt{HMx} in solid line, and \texttt{Horizon-Large} at $z=0$ in dotted line. The power spectra go from $z=0$ in dark blue to $z=4.25$ in yellow.}
    \label{fig:pressure_ps}
\end{figure*}

We first compare the pressure auto-power spectrum measured in the \texttt{Horizon-AGN} and \texttt{Magneticum} simulations to the one predicted by \texttt{HMx} as a function of redshift in Fig.~\ref{fig:pressure_ps}. Additionally, the result of \texttt{Horizon-noAGN} and \texttt{Horizon-Large} at $z \sim 0$ are included to emphasize the impact of different physics. We do not show their evolution with redshift since the trend is similar to the other simulations. At $z \sim 0$, both \texttt{Horizon-AGN} and \texttt{Horizon-noAGN} predict an excess of power, even more important in \texttt{Horizon-AGN}. On the other hand, \texttt{Horizon-Large} shows a power deficit. Finally, \texttt{Magneticum} is in relatively good agreement with the prediction at every scale. We also observe that there is more power in the larger simulations, which is expected as they contain more massive halos (see also Sect.~\ref{section:angulappshalomodel}). 

Then, we can examine the evolution with redshift. For all the simulations, we observe a relatively good agreement at low redshift with the predictions from \texttt{HMx} (up to $z \sim 1$ for \texttt{Horizon-AGN} and \texttt{Magneticum}). However, as the redshift increases, discrepancies become more evident. We observed that the differences in the pressure power spectrum are not attributed to the one of the matter power spectra, suggesting that these discrepancies are more likely due to the modelling of the pressure rather than the matter. \texttt{HMx} predicts an excess of power at high redshift, indicating that the model's physics fails to capture the nuances present in the simulations. We also notice that the measured power spectra are flatter than the predictions. Further investigations into these differences will be studied and discussed in Sect.~\ref{sec:discussion}.

\subsubsection{Matter-pressure power spectrum}
\begin{figure*}
    \centering
    \includegraphics[width=0.9\textwidth]{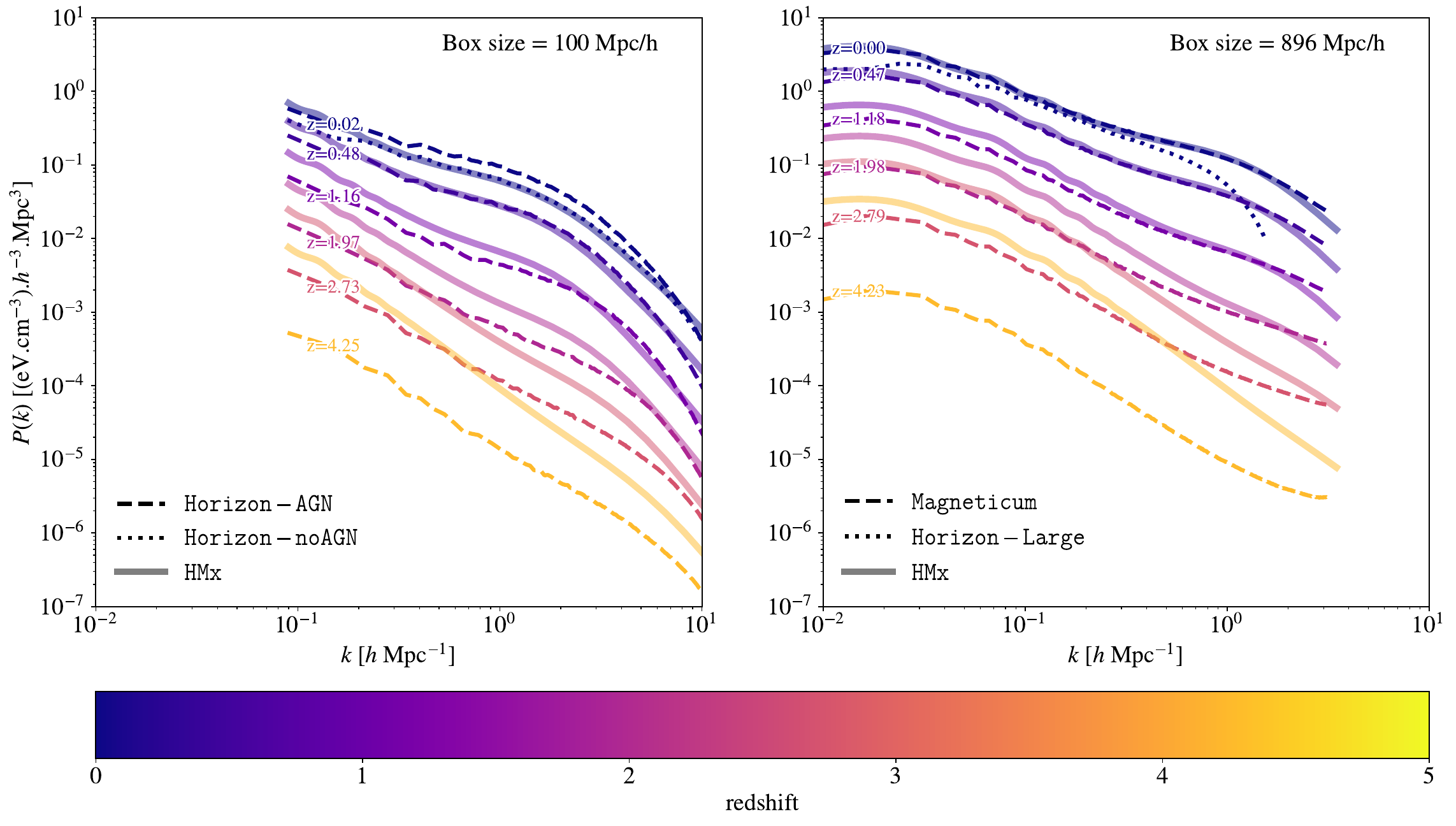}
    \caption{Matter-pressure power spectrum as a function of redshift. The left panel shows the results for simulations with a box size of $100\, h^{-1}\,\rm Mpc$, thus \texttt{Horizon-AGN} in dashed line compared to \texttt{HMx} in solid line, and \texttt{Horizon-noAGN} at $z=0$ in dotted line. The right panel shows the results for simulations with a box size of $896\, h^{-1}\,\rm Mpc$, thus \texttt{Magneticum} in dashed line compared to \texttt{HMx} in solid line, and \texttt{Horizon-Large} at $z=0$ in dotted line. The power spectra go from $z=0$ in dark blue to $z=4.25$ in yellow.}
    \label{fig:mxp}
\end{figure*}

As we are using the \texttt{HMx} model calibrated on the matter-pressure power spectrum (see Sect.~\ref{sec:hmx}), we extended our analysis to examine the matter-pressure power spectrum to explore the agreement across different redshifts. Moreover, modelling the matter-pressure power spectrum is important for studying the correlation between weak lensing and pressure (that we are not doing here). In Fig.~\ref{fig:mxp}, we show the measurement obtained from the \texttt{Horizon-AGN} and \texttt{Magneticum} simulations compared to the \texttt{HMx} prediction as a function of redshift. Additionally, the result of \texttt{Horizon-noAGN} and \texttt{Horizon-Large} at $z \sim 0$ are included to emphasize the impact of different physics. We do not show their evolution with redshift since the trend is similar to the other simulations. At $z \sim 0$, as for the pressure auto-power spectrum, \texttt{Horizon-AGN} demonstrates an excess of power, \texttt{Horizon-Large} a lack of power, while \texttt{Magneticum} agrees well with the prediction. We now observe a good agreement between \texttt{Horizon-noAGN} and \texttt{HMx}. 

Then, we can examine the evolution with redshift. We observe a better agreement up to a comparable redshift ($z\sim1.18$ for both simulations) than for the pressure auto-power spectrum. This outcome is expected as \texttt{HMx} has been calibrated on the matter-pressure power spectrum up to $z=1$. Moreover, the matter auto-power spectrum agrees well across all redshifts, mitigating the discrepancies in the pressure auto-power spectrum. At higher redshifts, the discrepancies caused by the pressure auto-power spectra persist. 

Given the sensitivity of pressure to baryonic physics, such cross-correlations can serve as valuable tools for constraining astrophysical parameters (e.g. with the cosmic shear-tSZ cross-correlation with the \texttt{flamingo} simulations in \citealp{McCarthy_2023} or with the lensing-tSZ cross-correlation from KiDS-1000 (lensing), Planck and ACT (tSZ) in \citealp{Troster_2022}). Depending on the probes we are working with, it is crucial to adequately model different redshift ranges. With future surveys, we expect to be sensitive up to redshift two for probes such as the distribution of galaxies, tomographic studies, or weak lensing (e.g. with Euclid, \citealp{Euclid_redbook} or Roman, \citealp{Eifler_2024}), our prediction thus needs to be trustable up to this redshift, which is qualitatively the case for the matter-pressure power spectrum.

\subsection{Angular power spectrum}
The observable accessed through surveys is the angular power spectrum, and its accurate prediction is crucial. Using the power spectra computed on the simulations or predicted by \texttt{HMx}, we used Eqs.~\eqref{eq:Cl} and \eqref{eq:Xy} to obtain the pressure angular power spectrum. We integrated these spectra over the redshift range $z = 0.02$ to $z = 4$ (according to the discussion in Sect.~\ref{section:angulappshalomodel}) and we limited our analysis to the Nyquist frequency: $k_{\rm Ny}\sim 16\, h^{-1}\, \rm Mpc$ for \texttt{Horizon-AGN} and \texttt{Horizon-noAGN}, while going up to $k_{\rm Ny} \sim 3.6 \, h^{-1}\, \rm Mpc$ for \texttt{Magneticum} and \texttt{Horizon-Large} (using a projection on $1024^3$ to achieve a comparable $k_{\rm Ny}$ to that of \texttt{Magneticum}). At each redshift, the $\ell$ range accessible with the simulations varies, depending on the available $k$ range (which is influenced by the size of the simulations). Choosing an $\ell$ range accessible to all the simulations across all redshifts between $0.02$ and $4$ would be quite narrow. To avoid this limitation and extract maximum information from the simulations, we opted for an interesting range of $\ell$, filling the angular power spectrum with a zero for any inaccessible $\ell$. We ensured to maintain the same behaviour in the angular power spectrum computed with the power spectrum of \texttt{HMx}, by cutting out at the same locations. We decided to compute the angular power spectrum for $\ell$ between $10^3$ and $10^4$ to encompass a range where the tSZ becomes an important foreground but remains accessible with future surveys. This pipeline limits our predictive power but allows us to infer the properties and behaviour of the different simulations and \texttt{HMx}.  

\begin{figure*}
    \centering
    \includegraphics[width=0.9\textwidth]{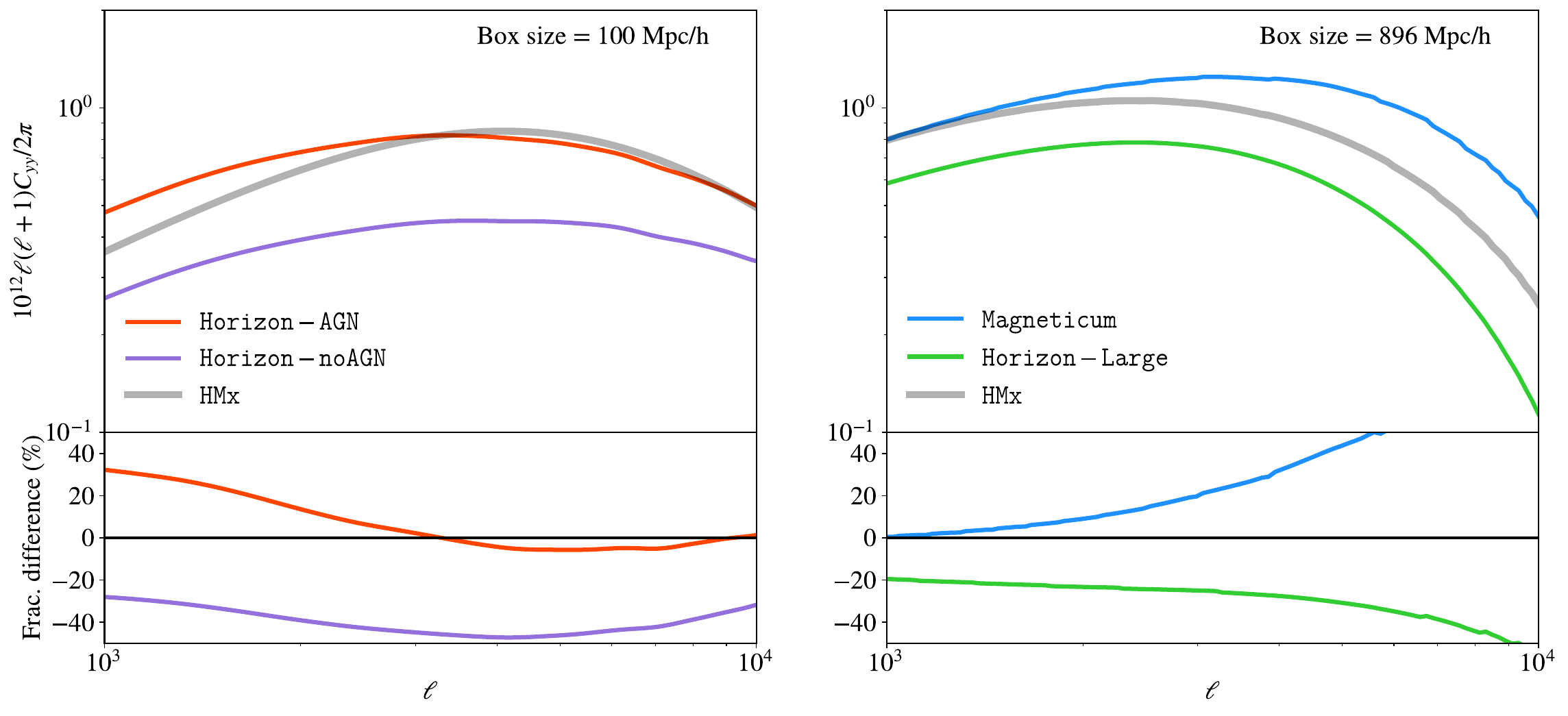}
    \caption{Pressure angular power spectrum integrated between $z=0$ and $z=4$ for the different simulations in different colours and \texttt{HMx} in dark grey. The left panel shows the results for simulation with a box size of $100\, h^{-1}\,\rm Mpc$, thus \texttt{Horizon-AGN} in red and \texttt{Horizon-noAGN} in purple. The right panel shows the results for simulation with a box size of $896\, h^{-1}\,\rm Mpc$, thus \texttt{Horizon-Large} in green and \texttt{Magneticum} in blue.} 
    \label{fig:Cl_yy}
\end{figure*}

The angular power spectra obtained from the different simulations are shown in Fig.~\ref{fig:Cl_yy} and compared with the \texttt{HMx} prediction. The agreement between \texttt{Horizon-AGN} and \texttt{HMx} improves at higher $\ell$ with differences reaching less than $10\,\%$ between $\ell = 3\times 10^3$ and $\ell = 10^4$. The power spectrum of \texttt{Horizon-AGN} (see Fig.~\ref{fig:pressure_ps}), exhibits more power at low redshifts and less power at high redshifts compared to the prediction, and these discrepancies seem to compensate each other, resulting in a small difference in the angular power spectrum. In contrast, \texttt{Magneticum} shows an opposite trend, with better agreement at low $\ell$ where differences are less than $10\,\%$ difference between $\ell = 10^3$ and $\ell = 2-3 \times 10^3$. For \texttt{Horizon-noAGN} and \texttt{Horizon-Large}, the general behaviour is quite similar, the simulations always have between $20\,\%$ and $50\,\%$ less power than \texttt{HMx}. 

\section{Discussion} \label{sec:discussion}
In the previous section, we have emphasized the observed differences between prediction from \texttt{HMx} and measurement from the simulations (\texttt{Horizon-AGN}, \texttt{Horizon-noAGN}, \texttt{Horizon-Large} and \texttt{Magneticum}), in particular, the increased discrepancy when the redshift increases. In this section, we explore some effects that can partially explain the observed differences between prediction and measurement but also between the measurements in different simulations. It can give a hint at the intrinsic limitation of a halo model but also an avenue for improved prediction. We are thus exploring the consequences of using a halo model prescription in Sect.~\ref{sec:halo_model_consequences}, the electron pressure profile in Sect.~\ref{sec:pressure_profile}, and the differences between simulations in Sect.~\ref{sec:diff_sim}.

\subsection{Halo model consequences} \label{sec:halo_model_consequences}
In this section, we are interested in studying the limitations caused by a halo model prescription.  We explored the one- and two-halo decomposition, the validity of the bound gas assumption at different redshifts, and the importance of the different masses. This can help us understand the limitations that can bias or limit our predictive power.

\subsubsection{One- and two-halo terms decomposition} \label{sec:1h-2h}
\begin{figure*}
    \centering
    \includegraphics[width=0.9\textwidth]{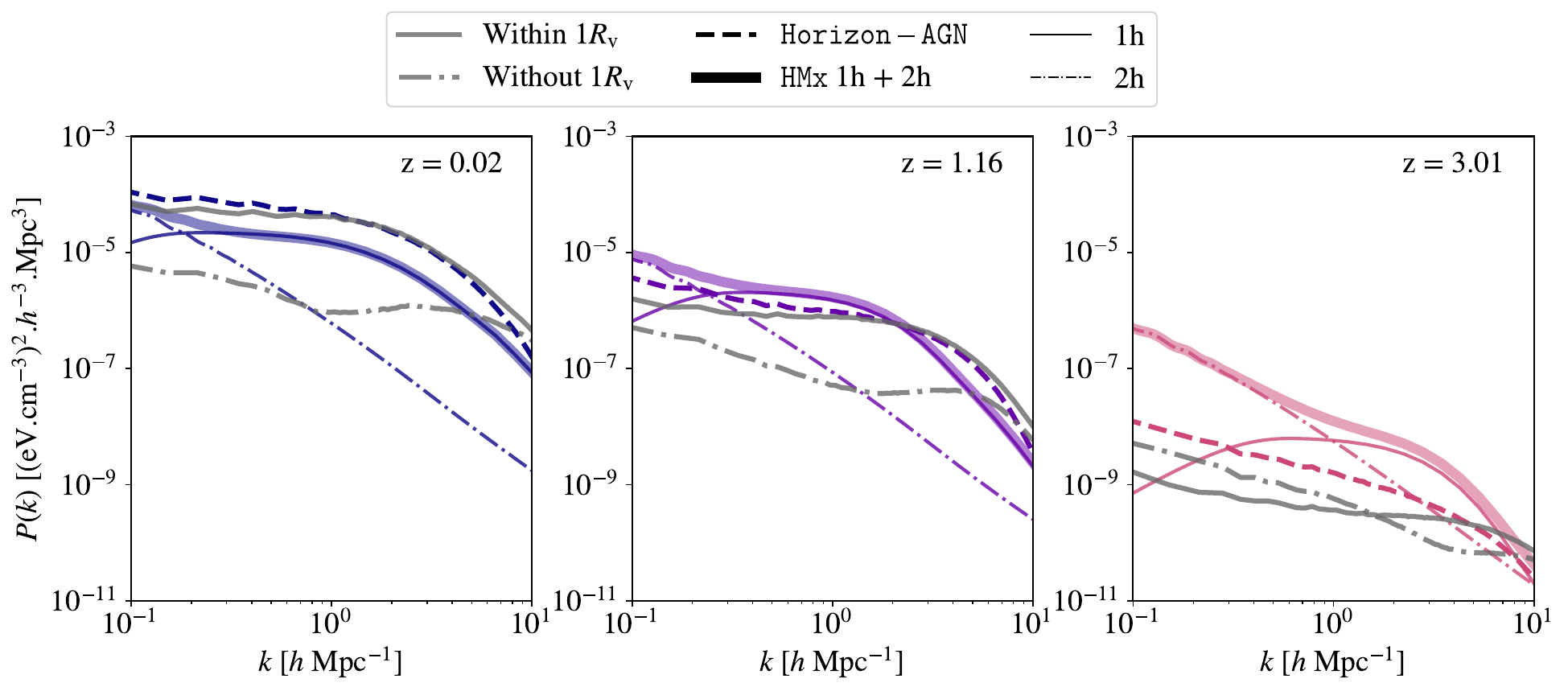}
    \caption{Pressure auto-power spectrum as a function of redshift in the different panels. For every redshift, we show the power spectrum of \texttt{Horizon-AGN} in a coloured dashed line, the \texttt{HMx} prediction in a coloured full line, the predicted one-halo term in a coloured thin line, and the predicted two-halo term in a coloured dotted-dashed line. We superimpose the pressure auto-power spectrum within (without) one virial radius in a grey full line (grey dotted-dashed line). On the left, we show the result for $z=0.02$, on the middle for $z=1.16$, and on the right for $z=3.01$.}
    \label{fig:1h-2h_Rvir}
\end{figure*}

To understand better the differences observed in the predicted and measured pressure auto-power spectrum (see Fig.~\ref{fig:pressure_ps} and related text), we explored the evolution of the one- and two-halo term contributions to the total power spectrum as a function of redshift. In Fig.~\ref{fig:1h-2h_Rvir}, we show this decomposition for three different redshifts: $z = 0.02$, $z = 1.16$, and $z=3.01$. For each redshift, we compared the \texttt{Horizon-AGN} power spectrum, the one predicted by \texttt{HMx}, and the one-halo and two-halo term predicted by \texttt{HMx}. As we increase the redshift, the contribution from the two-halo term becomes increasingly significant compared to the one-halo term at a given $k$. The increasing importance of the two-halo term is related to the scale at which the one- and two-halo terms intersect, which shifts towards higher $k$ values. We observe that the excess of power in the \texttt{HMx} power spectrum at higher redshifts is thus dominated by the two-halo term at low $k$ and by the one-halo term at high $k$. The excess of power in \texttt{HMx} suggests potential discrepancies in the distribution between the one- and two-halo terms, including the amplitude of each term at a given $k$ scale as a function of redshift. At higher redshifts, there are only a few halos, the observable thus resembles the matter distribution, whereas at lower redshifts, more, and more massive, halos have formed. Thus, at low redshifts, the hypothesis that all the matter is within the halos is more accurate and the halos contribute more to the total power. Also, the contribution from the halos increases more rapidly than that from the diffuse gas. Finally, differences can suggest an inaccurate representation of the intergalactic medium effects by the two-halo term, in particular its amplitude. 

\subsubsection{Validity of the bound gas assumption}
The halo model contains some intrinsic limitations, as it assumes that all the matter is within spherical halos and that the one-halo term -- which is the dominant contribution to the power spectra at low redshift (as shown by the coloured lines in Fig.~\ref{fig:1h-2h_Rvir} and discussed in the previous subsection) -- only contains bound gas, which is gas inside the virial radius. To better understand the validity of this assumption, we investigated how the power spectrum differs when considering pressure inside or outside one virial radius of halos, compared to the prediction from \texttt{HMx}. These spectra are obtained by selectively masking gas pressure by masking either inside or outside one virial radius of halos. In Fig.~\ref{fig:1h-2h_Rvir}, we show the contribution of bound and diffuse gas to the power spectrum measured from \texttt{Horizon-AGN}, and we find that the remaining simulations show similar trends. We compared the electron pressure auto-power spectra of \texttt{Horizon-AGN} to the one coming from inside (outside) one virial radius for three different redshifts: $z=0.02$, $z=1.16$, and $z=3.01$. We observe that, at low redshift, most of the power comes from within one virial radius of halos. However, this assumption loses validity with increasing redshifts. Notably, at $z=3.01$, there is more power coming from outside one virial radius of halos than inside. This implies the diminishing applicability of the halo model prescription at higher redshifts. Moreover, comparing the one-halo term to the power spectrum coming from inside one virial radius, reveals that at $z=0.02$, the shape is consistent even if \texttt{HMx} lacks power. However, as redshift increases, discrepancies in shape emerge, contrary to the expectation that the one-halo term should be representative of the power spectrum within one virial radius of the halos. These observations can partially explain the discrepancies observed in Fig.~\ref{fig:pressure_ps} with increasing redshift. Given these limitations, it becomes imperative to consider them when evaluating the cross-correlation with other probes.

\subsubsection{Importance of the halo mass}
\begin{figure}
    \centering
    \includegraphics[width=\hsize]{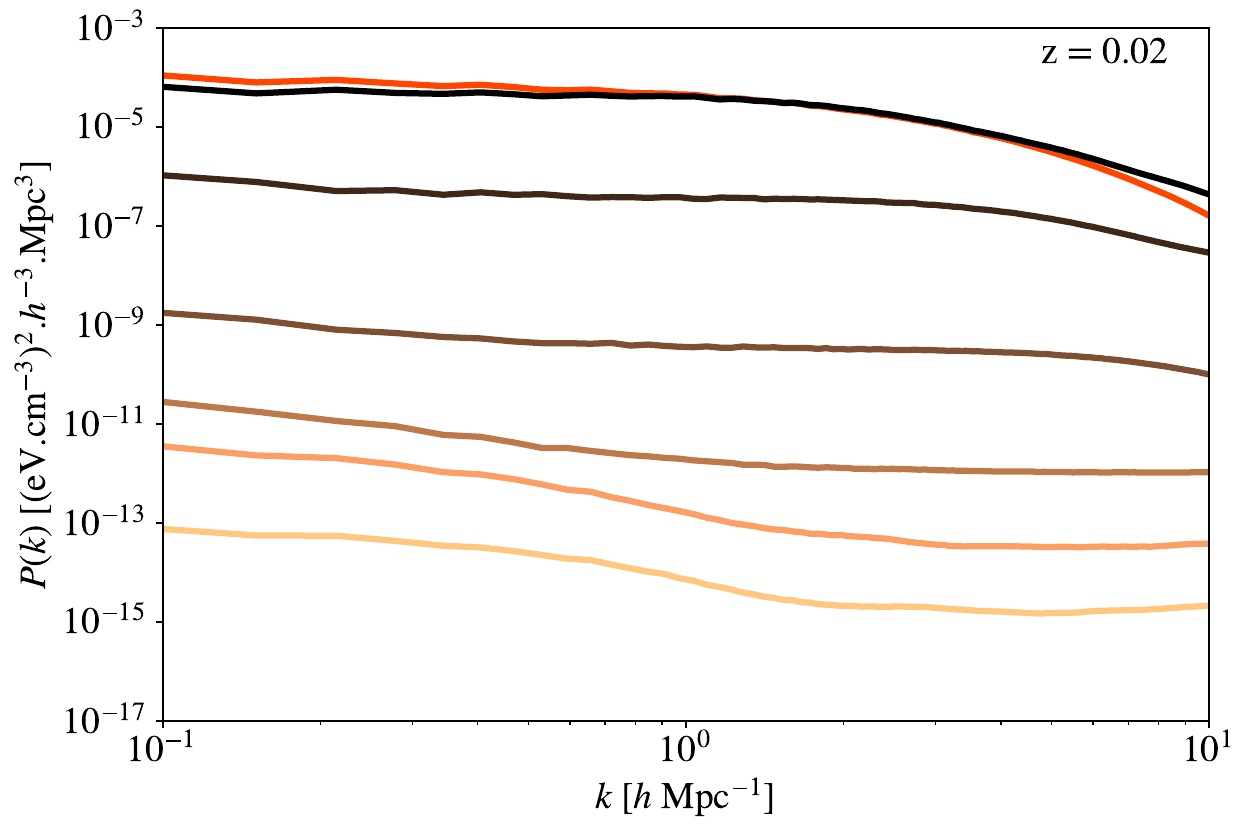}
    \caption{Pressure auto-power spectrum of the total \texttt{Horizon-AGN} simulation in red compared to the signal coming from within one virial radius of different mass bin. The highest mass bin is in dark brown and corresponds to the mass bin $14<\log (M_{\rm v}/M_\odot h^{-1})<15$, we then decrease of one dex for every curve until reaching the lowest mass bin in yellow, which corresponds to the mass bin $\log (M_{\rm v}/M_\odot h^{-1})<10$.}
    \label{fig:xRvir}
\end{figure}

To go one step further in our test of the halo model, we investigated the contribution of each mass bin to the overall power spectrum. We performed this study for the different simulations (\texttt{Horizon-AGN}, \texttt{Horizon-noAGN}, \texttt{Horizon-Large} and \texttt{Magneticum}), and redshifts, and they all show a similar trend, thus, we are only presenting and quantify the results for \texttt{Horizon-AGN} at $z=0.02$ in Fig.~\ref{fig:xRvir}. We compared the electron pressure auto-power spectra with the one coming from inside one virial radius of the halos in different mass bins at $z=0.02$. We can clearly see that most of the power emanates from the highest mass bin, despite its relatively low population (for \texttt{Horizon-AGN}, the highest mass bin contains only $0.006\,\%$ of the total number of halos). The lower the mass bin, the lesser its contribution to the total power spectrum. A reduction of approximately one order of magnitude is observed with each logarithmic decrease in mass bin. This shows the importance of ensuring that the halo model, particularly the electron pressure profile, accurately reflects the characteristics of the highest mass halos, in agreement with the conclusion done in Sect.~\ref{section:angulappshalomodel}. However, given that the higher mass halos are less common occurrences, it may be worth masking them to mitigate the connected non-Gaussian covariance and tighten cosmological or astrophysical constraints, as explored in \cite{Osato_2021}. 

\subsection{Halo pressure profiles} \label{sec:pressure_profile}
\begin{figure*}[!t]
    \centering
    \includegraphics[width=0.95\textwidth]{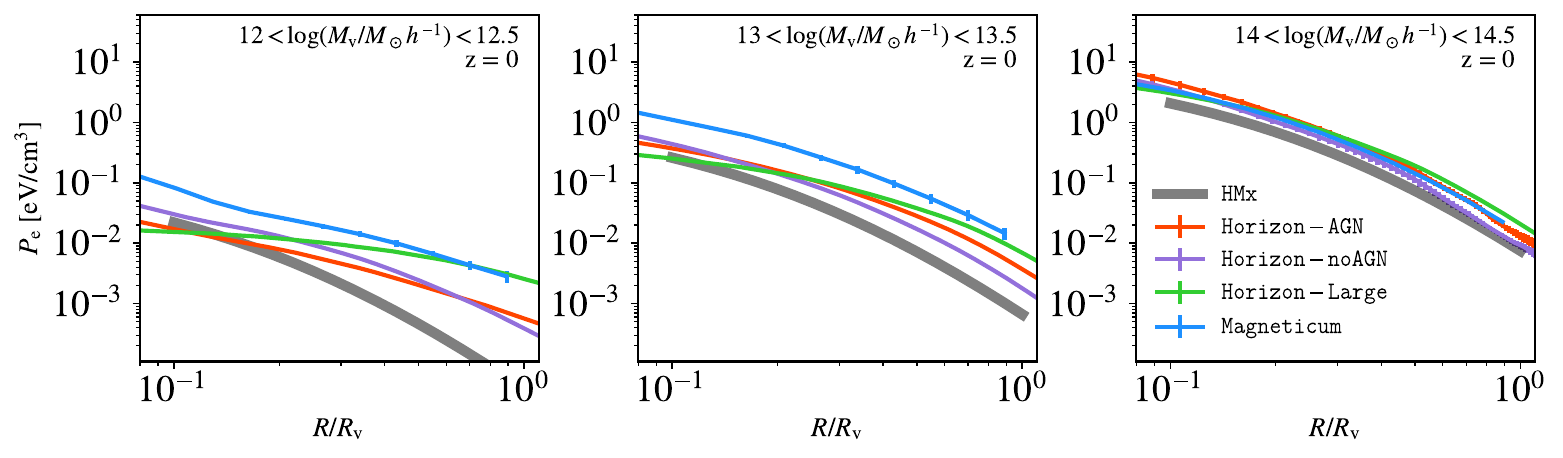}
    \includegraphics[width=0.95\textwidth]{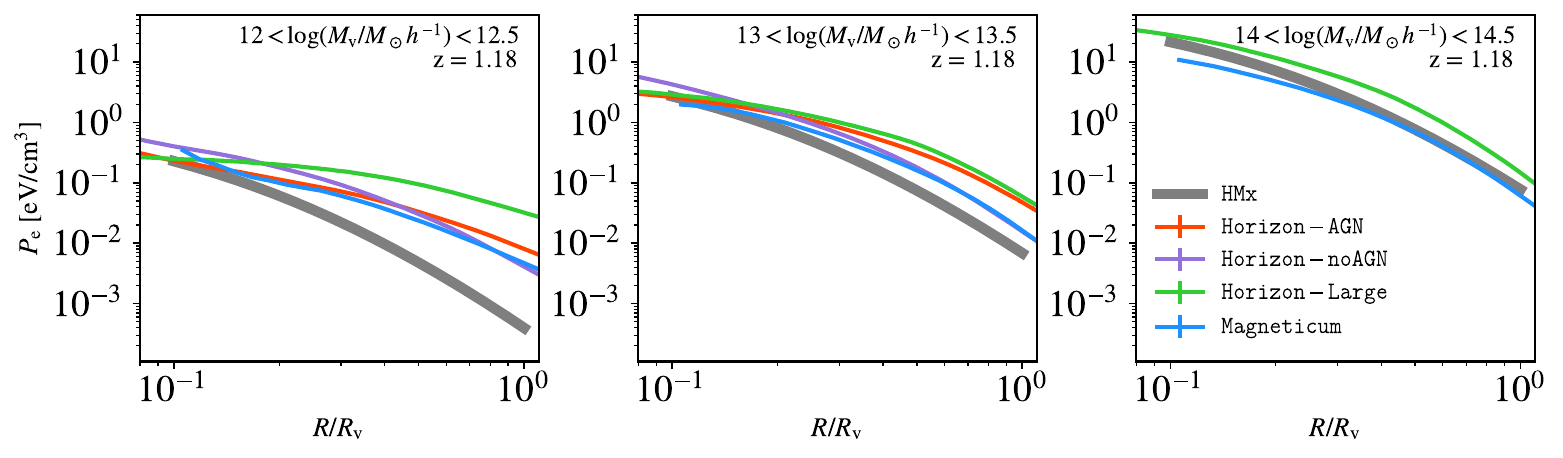}
    \caption{Pressure profile as a function of the distance to the centre of the halos compared with the one predicted by \texttt{HMx} in dark grey at $z=0$ (first row) and $z=1.18$ (second row). Each panel represents a different mass bin and each colour is a different simulation: \texttt{Horizon-AGN} in red, \texttt{Horizon-noAGN} in purple, \texttt{Horizon-Large} in green, and \texttt{Magneticum} in blue. The error bars represent the error on the mean and are most of the time too small to be visible.}
    \label{fig:profile_z0}
\end{figure*}

One of the main components of the model is the electron pressure profile, defined in Eq.~\eqref{eq:profile}. This profile not only represents a major aspect of the model but also constitutes the primary component of the one-halo term. As demonstrated in Fig.~\ref{fig:1h-2h_Rvir}, the one-halo term is the predominant contribution to the power spectrum at low redshifts, further influencing the angular power spectrum. Consequently, in Fig.~\ref{fig:profile_z0} we present the electron pressure profile measured in the different simulations (\texttt{Horizon-AGN}, \texttt{Horizon-noAGN}, \texttt{Horizon-Large}, and \texttt{Magneticum}) compared to the prediction from \texttt{HMx} at $z=0$ and $z=1.18$, across three mass bins: $12 < \log (M_{\rm v}/M_\odot h^{-1}) < 12.5$, $13 < \log (M_{\rm v}/M_\odot h^{-1}) < 13.5$ and $14 < \log (M_{\rm v}/M_\odot h^{-1}) < 14.5$. The consistency in trends between the two redshifts aligns with expectations, given that the measured profiles scale as $(1+z)^4$, as the predicted one. For all the \texttt{Horizon} simulations at both redshifts, as well as \texttt{Magneticum} at $z=1.18$, discrepancies are noticeable, particularly in the low- and intermediate-mass bins, where the inner regions exhibit closer agreement, while deviations escalate towards outer regions. The \texttt{Magneticum} profiles at $z=0$ have different behaviour, characterised by elevated inner pressures and convergence towards other simulations in the outer regions. For the high-mass bin, we see that the \texttt{Horizon} simulations have a higher amplitude than the predictions at both redshifts, whereas the \texttt{Magnetiucm} profiles surpass the prediction at $z=0$ and undershoot them at $z=1.18$. Nevertheless, the overall shape remains reasonably consistent across all distances from the centre. As \texttt{HMx} should represent the mean behaviour of halos, we have added error bars that indicate the error on the mean, and are often too small to be discernible. This observation suggests that while our mean is well estimated, it is not entirely compatible with the \texttt{HMx} prescription.

Since the model's free parameters are tailored to fit the response power spectrum, achieving a perfect agreement on pressure profiles is not guaranteed. Moreover, the influence of high masses, which contribute the most to the power spectrum, can alter even more the profiles of low-mass halos that contribute slightly to the power spectrum. The predicted profiles should be considered as the ones of effective halos, but it is still worth comparing the predicted and measured profiles, to understand power spectrum disparities. Let us describe more the high-mass bin which is the dominant part of the power spectrum. The difference observed in this bin could explain power spectrum differences. For example, at $z=0$, all simulations (\texttt{Horizon-AGN}, \texttt{Horizon-noAGN}, \texttt{Horizon-Large} and \texttt{Magneticum}) have qualitatively similar pressure profiles (with more pronounced differences on the outer regions), containing more pressure than \texttt{HMx}. At $z=1.18$, the \texttt{Horizon-Large} remains above the prediction, while \texttt{Magneticum} is under. On the left panel of Fig.~\ref{fig:pressure_ps}, both \texttt{Horizon-AGN} and \texttt{Horizon-noAGN} power spectra also contain more power than \texttt{HMx}. However, on the right panel of Fig.~\ref{fig:pressure_ps}, \texttt{Horizon-Large} power spectrum lies below \texttt{HMx} (with \texttt{Magneticum} showing relatively good agreement). These diverse behaviours suggest that profile differences alone cannot entirely account for observed power spectrum discrepancies. Additionally, we note that the lower-mass halos exhibit greater discrepancies, as anticipated.

To improve our analysis, it can be interesting to focus even more on high-mass halos and their impact on the tSZ properties. Future studies with larger volume simulations could provide a more comprehensive probe of these halos, which are quite rare in the \texttt{Horizon} suite, for example. Because of the current computational constraints, the resolution of such simulations cannot be as good as the one in \texttt{Horizon-AGN}, which can potentially introduce additional biases. Building large simulations with zoom-in capabilities targeting big halos to assess the fidelity of baryonic physics can be an avenue. This approach could offer a new perspective for such analysis, which is currently limited by the noise on the number of these halos. 

\subsection{Difference of the simulations} \label{sec:diff_sim}
The simulations that we are analysing are different: they are run with different computational codes, different physics models, different resolutions, and different box sizes. Different choices in terms of the included physics and their modelisation methodology are made and can influence the obtained results. For example, in \cite{Mead_2020}, models were fitted against three \texttt{BAHAMAS} simulations, each containing different strengths of AGN feedback, yielding different values for the fitted parameters. In our simulations, with the different choices, the strength of feedback is also different, but other choices can lead to many other differences that are challenging to precisely identify and define. 

Another critical aspect that can be evaluated is the cosmic variance of the simulation. To quantity this variance, we took 500 non-overlapping boxes extracted from the \texttt{Horizon-Large} simulation, each with a dimension of $100 \,h^{-1}\, \rm Mpc$. As the power spectrum depends a lot on the high-mass halos, we applied selection criteria to retain only boxes with maximal mass similar to the one in the \texttt{Horizon-AGN} and \texttt{Horizon-noAGN} simulations. This refinement yielded $363$ non-overlapping boxes, from which we computed the pressure auto-power spectrum. We employed a constant binning in $\log$ space of $k$ for our power spectrum, checking that the level of correlation between the bins is low. We then calculated a cumulative probability distribution function at each $k$ to extract the values encompassing $68\%$ of the signal. This approach allowed us to derive the lower and upper bounds of our error bars. 

\begin{figure}[!t]
    \centering
    \includegraphics[width=0.95\hsize]{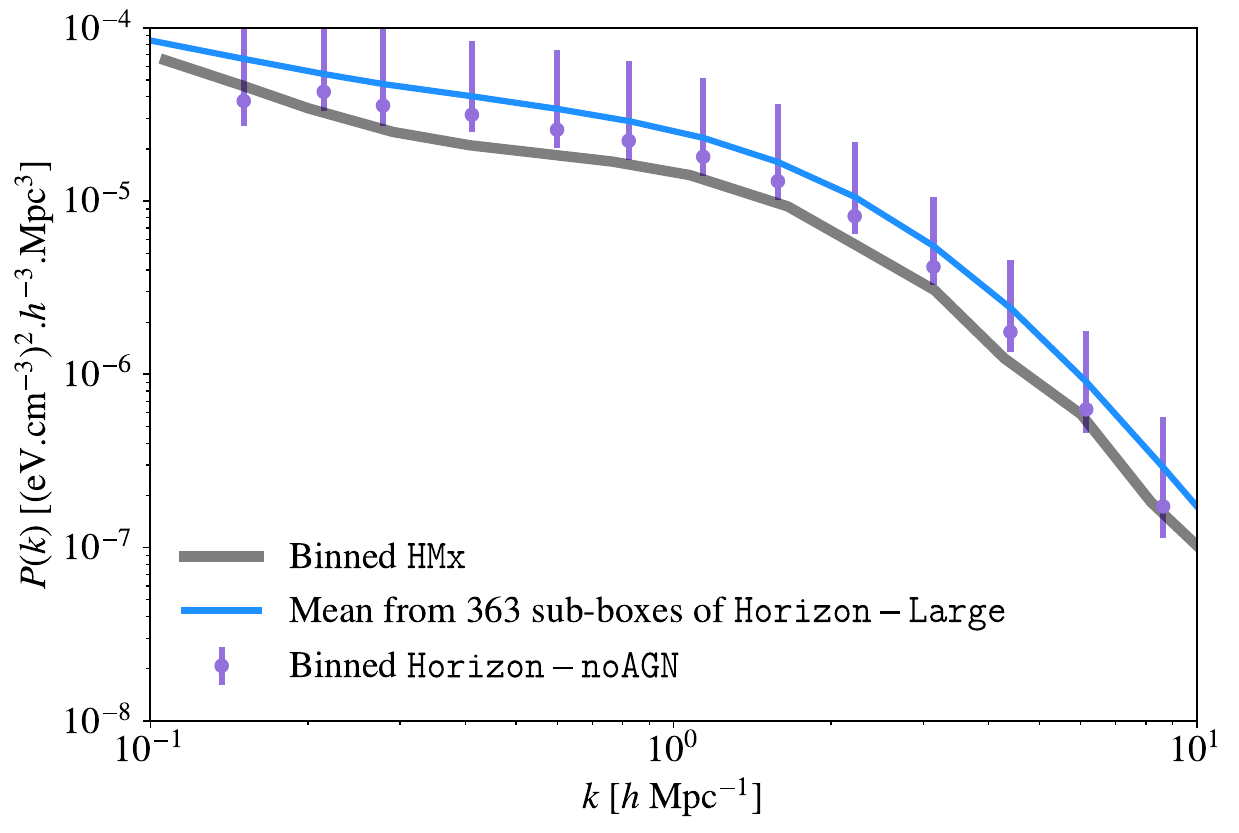}
    \caption{Pressure auto-power spectrum in \texttt{Horizon-noAGN} in purple compared to \texttt{HMx} in dark grey and the mean over 363 sub-boxes of $100\,h^{-1}\,\rm Mpc$ from the \texttt{Horizon-Large} in blue. The purple error bars contain $68\%$ around the mean of the 363 sub-boxes of the \texttt{Horizon-Large} (see the text for more information). They are applied to the \texttt{Horizon-noAGN} simulations to represent its variance.}
    \label{fig:variance}
\end{figure}

In Fig.~\ref{fig:variance} we show the binned mean power spectrum of these boxes, of the \texttt{Horizon-noAGN} simulation, and of the \texttt{HMx} prediction. The variance derived using the method described above is overlaid on the \texttt{Horizon-noAGN} simulation to represent the cosmic variance of such a simulation. We see that the error bars are non-Gaussian and of the same order of magnitude across different $k$ ranges. They emphasize a tendency for a higher power spectrum than the one predicted by \texttt{HMx}.

\section{Conclusions}
\label{conclu}

Within this paper, we compared the predicted (angular) power spectrum from \texttt{HMx} to the measurement from different simulations (\texttt{Horizon-AGN}, \texttt{Horizon-noAGN}, \texttt{Horizon-Large} and \texttt{Magneticum}). Using the \texttt{HMx} halo model, we predicted the pressure angular power spectra by integrating up to different redshifts or different masses. Our analysis first reveal that integration up to $z=4$ (or even $z=3$) accounts for more than $98\,\%$ ($96\,\%$) of the power. We also find that integrating up to $4\times 10^{15}\, h^{-1} \, M_\odot$ captures $97\,\%$ of the signal, highlighting the significant contribution of higher masses to the power (this is also evident in Fig.~\ref{fig:xRvir} where we examined the power contribution from different mass bins at the level of the power spectrum). From these initial findings, we conclude that our comparison should focus on the redshift range between $0$ and $4$ and emphasize the importance of the highest mass range.

The main results of our analysis can be summarised in three points:
\begin{itemize}
    \item The comparison between the pressure power spectrum from different simulations with the one predicted by \texttt{HMx} for redshifts between $z=0$ and $z=4.25$ (Fig.~\ref{fig:pressure_ps}), reveals a qualitative conclusion which is consistent across all simulations: at low redshift, there is a qualitative agreement between measurements and predictions but discrepancies increase with higher redshifts.
    \item A similar trend is observed for the matter-pressure power spectrum (Fig.~\ref{fig:mxp}), although the discrepancies are less pronounced due to better agreement in the matter-auto power spectra. It is important to note that when working with cross-correlation such as the weak lensing-tSZ one, sensitivity extends only up to a redshift $\sim 2$. This range is where \texttt{HMx} shows relatively good agreement.
    \item The comparison between the predicted and measured pressure angular power spectrum reveals differences ranging from $20$ to $50$\% across all simulations. 
\end{itemize}

To understand the origin of the discrepancies, we explored the limitations inherent to the halo model:
\begin{itemize}
    \item we observe that the excess of power in the \texttt{HMx} prediction is dominated by the one-halo term at high $k$ and by the two-halo term at low $k$ (Fig.~\ref{fig:1h-2h_Rvir}),
    \item we investigated the contribution of power from within and outside one virial radius of halos in the simulations at different redshifts (Fig.~\ref{fig:1h-2h_Rvir}). At lower redshifts, the majority of power originates from within halos, indicating compatibility with the halo model assumption. However, at higher redshift (e.g. $z\sim 3$), a substantial portion of the power comes from regions outside one virial radius of halos, highlighting a limitation of the halo model to capture this phenomenon,
    \item we compared the pressure profiles across different mass bins and redshifts, as the pressure profile is the main property of the halo model (Fig.~\ref{fig:profile_z0}). While differences were expected due to the halo model being fitted at the level of the response power spectrum, it was informative to observe the behaviour and degeneracies. Across most mass and redshift ranges, the simulations exhibit higher power levels and distinct profile shapes compared to the predictions. 
\end{itemize}

Enhancing the robustness of predictions could be achieved by developing a halo model that more accurately reflects the physics within simulations.  This could involve calibrations on simulations at the level of individual properties, such as the pressure profile. Additionally, to apply a halo model framework in cosmological analyses, calibrating the model with real data would be beneficial. Such approaches come with inherent challenges, such as the parameter measurement methodologies. Exploring different parametrisations that better align with these properties may help determine whether an accurate match at the component level would lead to a robust power spectrum prediction. Employing larger and better-resolved simulations, which include more massive halos, can also address some limitations of the halo model, particularly by enhancing our understanding of the tSZ effect and the behaviour of high-mass halos across different redshifts. Another significant avenue for exploration is investigating the variability of the tSZ power spectrum under different cosmologies. This includes assessing the influence of cosmological parameters that impact the growth of structures, such as the quantity of baryonic matter or the dark energy equation of state. 

\begin{acknowledgements}

We thank Klaus Dolag for giving us access to the Magneticum simulations and the useful discussions. This work has made use of the Infinity Cluster hosted by Institut d'Astrophysique de Paris. We thank St\'ephane Rouberol for running smoothly this cluster for us. 

EA, KB, and YD acknowledge support by the CNES, the French National Space Agency. EA and PRS are supported in part by a PhD Joint Program between the CNRS and the University of Arizona. EK is supported in part by Department of Energy grant DE-SC0020247, the David and Lucile Packard Foundation and an Alfred P. Sloan Research Fellowship. 
The contributions of the authors, classified according to the CRediT (Contribution Roles Taxonomy)\footnote{\href{https://authorservices.wiley.com/author-resources/Journal-Authors/open-access/credit.html}{https://authorservices.wiley.com/author-resources/Journal-Authors/open-access/credit.html}} system, were as follows: Emma Ayçoberry: Conceptualization; Formal Analysis; Software; Validation; Visualization; Writing – Original Draft Preparation. Pranjal R. S.: Formal Analysis; Validation; Visualization; Writing – Review \& Editing. Karim Benabed: Conceptualization; Validation; Writing – Original Draft Preparation. Yohan Dubois: Conceptualization; Data Curation; Software; Validation; Writing – Original Draft Preparation. Elisabeth Krause: Conceptualization; Validation; Writing – Review \& Editing. Tim Eifler: Validation; Writing – Review \& Editing.
\end{acknowledgements}

\newpage

\bibliographystyle{aa} 
\bibliography{bibli} 

\begin{thebibliography}{74}
\expandafter\ifx\csname natexlab\endcsname\relax\def\natexlab#1{#1}\fi

\bibitem[{{Aubert} {et~al.}(2004){Aubert}, {Pichon}, \&
  {Colombi}}]{Aubert_2004}
{Aubert}, D., {Pichon}, C., \& {Colombi}, S. 2004, \mnras, 352, 376

\bibitem[{{Battaglia} {et~al.}(2012){Battaglia}, {Bond}, {Pfrommer}, \&
  {Sievers}}]{Battaglia_2012}
{Battaglia}, N., {Bond}, J.~R., {Pfrommer}, C., \& {Sievers}, J.~L. 2012, \apj,
  758, 75

\bibitem[{{Beckmann} {et~al.}(2017){Beckmann}, {Devriendt}, {Slyz}, {Peirani},
  {Richardson}, {Dubois}, {Pichon}, {Chisari}, {Kaviraj}, {Laigle}, \&
  {Volonteri}}]{Beckmann17}
{Beckmann}, R.~S., {Devriendt}, J., {Slyz}, A., {et~al.} 2017, \mnras, 472, 949

\bibitem[{{Bleem} {et~al.}(2022){Bleem}, {Crawford}, {Ansarinejad}, {Benson},
  {Bocquet}, {Carlstrom}, {Chang}, {Chown}, {Crites}, {de Haan}, {Dobbs},
  {Everett}, {George}, {Gualtieri}, {Halverson}, {Holder}, {Holzapfel},
  {Hrubes}, {Knox}, {Lee}, {Luong-Van}, {Marrone}, {McMahon}, {Meyer},
  {Millea}, {Mocanu}, {Mohr}, {Natoli}, {Omori}, {Padin}, {Pryke},
  {Raghunathan}, {Reichardt}, {Ruhl}, {Schaffer}, {Shirokoff}, {Staniszewski},
  {Stark}, {Vieira}, \& {Williamson}}]{Bleem_2022}
{Bleem}, L.~E., {Crawford}, T.~M., {Ansarinejad}, B., {et~al.} 2022, \apjs,
  258, 36

\bibitem[{{Bolliet} {et~al.}(2018){Bolliet}, {Comis}, {Komatsu}, \&
  {Mac{\'\i}as-P{\'e}rez}}]{Bolliet_2018}
{Bolliet}, B., {Comis}, B., {Komatsu}, E., \& {Mac{\'\i}as-P{\'e}rez}, J.~F.
  2018, \mnras, 477, 4957

\bibitem[{{Bolliet} {et~al.}(2023){Bolliet}, {Kusiak}, {McCarthy}, {Sabyr},
  {Surrao}, {Hill}, {Chluba}, {Ferraro}, {Hadzhiyska}, {Han},
  {Mac{\'\i}as-P{\'e}rez}, {Madhavacheril}, {Maniyar}, {Mehta}, {Pandey},
  {Schaan}, {Sherwin}, {Spurio Mancini}, \& {Zubeldia}}]{Bolliet_2023}
{Bolliet}, B., {Kusiak}, A., {McCarthy}, F., {et~al.} 2023, arXiv e-prints,
  arXiv:2310.18482

\bibitem[{{Bryan} \& {Norman}(1998)}]{Bryan_1998}
{Bryan}, G.~L. \& {Norman}, M.~L. 1998, \apj, 495, 80

\bibitem[{{Chisari} {et~al.}(2018){Chisari}, {Richardson}, {Devriendt},
  {Dubois}, {Schneider}, {Le Brun}, {Beckmann}, {Peirani}, {Slyz}, \&
  {Pichon}}]{Chisari18}
{Chisari}, N.~E., {Richardson}, M.~L.~A., {Devriendt}, J., {et~al.} 2018,
  \mnras, 480, 3962

\bibitem[{{Davis} {et~al.}(2011){Davis}, {Devriendt}, {Colombi}, {Silk}, \&
  {Pichon}}]{Davis_2011}
{Davis}, O., {Devriendt}, J., {Colombi}, S., {Silk}, J., \& {Pichon}, C. 2011,
  \mnras, 413, 2087

\bibitem[{{Debackere} {et~al.}(2020){Debackere}, {Schaye}, \&
  {Hoekstra}}]{Debackere_2020}
{Debackere}, S. N.~B., {Schaye}, J., \& {Hoekstra}, H. 2020, \mnras, 492, 2285

\bibitem[{{Dolag} {et~al.}(2016){Dolag}, {Komatsu}, \& {Sunyaev}}]{Dolag_2016}
{Dolag}, K., {Komatsu}, E., \& {Sunyaev}, R. 2016, \mnras, 463, 1797

\bibitem[{{Dolag} \& {Stasyszyn}(2009)}]{Dolag_2009}
{Dolag}, K. \& {Stasyszyn}, F. 2009, \mnras, 398, 1678

\bibitem[{{Dubois} {et~al.}(2012){Dubois}, {Devriendt}, {Slyz}, \&
  {Teyssier}}]{Dubois12}
{Dubois}, Y., {Devriendt}, J., {Slyz}, A., \& {Teyssier}, R. 2012, \mnras, 420,
  2662

\bibitem[{{Dubois} {et~al.}(2016){Dubois}, {Peirani}, {Pichon}, {Devriendt},
  {Gavazzi}, {Welker}, \& {Volonteri}}]{Dubois_2016}
{Dubois}, Y., {Peirani}, S., {Pichon}, C., {et~al.} 2016, \mnras, 463, 3948

\bibitem[{{Dubois} {et~al.}(2014){Dubois}, {Pichon}, {Welker}, {Le Borgne},
  {Devriendt}, {Laigle}, {Codis}, {Pogosyan}, {Arnouts}, {Benabed}, {Bertin},
  {Blaizot}, {Bouchet}, {Cardoso}, {Colombi}, {de Lapparent}, {Desjacques},
  {Gavazzi}, {Kassin}, {Kimm}, {McCracken}, {Milliard}, {Peirani}, {Prunet},
  {Rouberol}, {Silk}, {Slyz}, {Sousbie}, {Teyssier}, {Tresse}, {Treyer},
  {Vibert}, \& {Volonteri}}]{Dubois_2014}
{Dubois}, Y., {Pichon}, C., {Welker}, C., {et~al.} 2014, \mnras, 444, 1453

\bibitem[{{Dubois} \& {Teyssier}(2008)}]{Dubois08}
{Dubois}, Y. \& {Teyssier}, R. 2008, \aap, 477, 79

\bibitem[{{Eifler} {et~al.}(2024){Eifler}, {Fang}, {Krause}, {Hirata}, {Xu},
  {Benabed}, {Ferraro}, {Miranda}, {S.}, {Ay{\c{c}}oberry}, \&
  {Dubois}}]{Eifler_2024}
{Eifler}, T., {Fang}, X., {Krause}, E., {et~al.} 2024, arXiv e-prints,
  arXiv:2411.04088

\bibitem[{{Fang} {et~al.}(2024){Fang}, {Krause}, {Eifler}, {Ferraro},
  {Benabed}, {Pranjal}, {Ay{\c{c}}oberry}, {Dubois}, \& {Miranda}}]{Fang_2023}
{Fang}, X., {Krause}, E., {Eifler}, T., {et~al.} 2024, \mnras, 527, 9581

\bibitem[{{Fedeli}(2014)}]{Fedeli_2014}
{Fedeli}, C. 2014, \jcap, 2014, 028

\bibitem[{{Ferragamo} {et~al.}(2023){Ferragamo}, {de Andres}, {Sbriglio},
  {Cui}, {De Petris}, {Yepes}, {Dupuis}, {Jarraya}, {Lahouli}, {De Luca},
  {Gianfagna}, \& {Rasia}}]{Ferragamo_2023}
{Ferragamo}, A., {de Andres}, D., {Sbriglio}, A., {et~al.} 2023, \mnras, 520,
  4000

\bibitem[{{Gonzalez} {et~al.}(2013){Gonzalez}, {Sivanandam}, {Zabludoff}, \&
  {Zaritsky}}]{Gonzalez_2013}
{Gonzalez}, A.~H., {Sivanandam}, S., {Zabludoff}, A.~I., \& {Zaritsky}, D.
  2013, \apj, 778, 14

\bibitem[{{Haardt} \& {Madau}(1996)}]{Haardt96}
{Haardt}, F. \& {Madau}, P. 1996, \apj, 461, 20

\bibitem[{{Hinshaw} {et~al.}(2013){Hinshaw}, {Larson}, {Komatsu}, {Spergel},
  {Bennett}, {Dunkley}, {Nolta}, {Halpern}, {Hill}, {Odegard}, {Page}, {Smith},
  {Weiland}, {Gold}, {Jarosik}, {Kogut}, {Limon}, {Meyer}, {Tucker}, {Wollack},
  \& {Wright}}]{Hinshaw_2013}
{Hinshaw}, G., {Larson}, D., {Komatsu}, E., {et~al.} 2013, \apjs, 208, 19

\bibitem[{{Hirschmann} {et~al.}(2014){Hirschmann}, {Dolag}, {Saro}, {Bachmann},
  {Borgani}, \& {Burkert}}]{Hirschmann_2014}
{Hirschmann}, M., {Dolag}, K., {Saro}, A., {et~al.} 2014, \mnras, 442, 2304

\bibitem[{{Holz} \& {Perlmutter}(2012)}]{Holz_2012}
{Holz}, D.~E. \& {Perlmutter}, S. 2012, \apjl, 755, L36

\bibitem[{{Komatsu} \& {Kitayama}(1999)}]{Komatsu_1999}
{Komatsu}, E. \& {Kitayama}, T. 1999, \apjl, 526, L1

\bibitem[{{Komatsu} \& {Seljak}(2001)}]{Komatsu_2001}
{Komatsu}, E. \& {Seljak}, U. 2001, \mnras, 327, 1353

\bibitem[{{Komatsu} \& {Seljak}(2002)}]{Komatsu_2002}
{Komatsu}, E. \& {Seljak}, U. 2002, \mnras, 336, 1256

\bibitem[{{Komatsu} {et~al.}(2011){Komatsu}, {Smith}, {Dunkley}, {Bennett},
  {Gold}, {Hinshaw}, {Jarosik}, {Larson}, {Nolta}, {Page}, {Spergel},
  {Halpern}, {Hill}, {Kogut}, {Limon}, {Meyer}, {Odegard}, {Tucker}, {Weiland},
  {Wollack}, \& {Wright}}]{Komatsu_2011}
{Komatsu}, E., {Smith}, K.~M., {Dunkley}, J., {et~al.} 2011, \apjs, 192, 18

\bibitem[{{Laureijs} {et~al.}(2011){Laureijs}, {Amiaux}, {Arduini},
  {Augu{\`e}res}, {Brinchmann}, {Cole}, {Cropper}, {Dabin}, {Duvet}, {Ealet},
  {Garilli}, {Gondoin}, {Guzzo}, {Hoar}, {Hoekstra}, {Holmes}, {Kitching},
  {Maciaszek}, {Mellier}, {Pasian}, {Percival}, {Rhodes}, {Saavedra Criado},
  {Sauvage}, {Scaramella}, {Valenziano}, {Warren}, {Bender}, {Castander},
  {Cimatti}, {Le F{\`e}vre}, {Kurki-Suonio}, {Levi}, {Lilje}, {Meylan},
  {Nichol}, {Pedersen}, {Popa}, {Rebolo Lopez}, {Rix}, {Rottgering},
  {Zeilinger}, {Grupp}, {Hudelot}, {Massey}, {Meneghetti}, {Miller}, {Paltani},
  {Paulin-Henriksson}, {Pires}, {Saxton}, {Schrabback}, {Seidel}, {Walsh},
  {Aghanim}, {Amendola}, {Bartlett}, {Baccigalupi}, {Beaulieu}, {Benabed},
  {Cuby}, {Elbaz}, {Fosalba}, {Gavazzi}, {Helmi}, {Hook}, {Irwin}, {Kneib},
  {Kunz}, {Mannucci}, {Moscardini}, {Tao}, {Teyssier}, {Weller}, {Zamorani},
  {Zapatero Osorio}, {Boulade}, {Foumond}, {Di Giorgio}, {Guttridge}, {James},
  {Kemp}, {Martignac}, {Spencer}, {Walton}, {Bl{\"u}mchen}, {Bonoli},
  {Bortoletto}, {Cerna}, {Corcione}, {Fabron}, {Jahnke}, {Ligori}, {Madrid},
  {Martin}, {Morgante}, {Pamplona}, {Prieto}, {Riva}, {Toledo}, {Trifoglio},
  {Zerbi}, {Abdalla}, {Douspis}, {Grenet}, {Borgani}, {Bouwens}, {Courbin},
  {Delouis}, {Dubath}, {Fontana}, {Frailis}, {Grazian}, {Koppenh{\"o}fer},
  {Mansutti}, {Melchior}, {Mignoli}, {Mohr}, {Neissner}, {Noddle}, {Poncet},
  {Scodeggio}, {Serrano}, {Shane}, {Starck}, {Surace}, {Taylor},
  {Verdoes-Kleijn}, {Vuerli}, {Williams}, {Zacchei}, {Altieri}, {Escudero
  Sanz}, {Kohley}, {Oosterbroek}, {Astier}, {Bacon}, {Bardelli}, {Baugh},
  {Bellagamba}, {Benoist}, {Bianchi}, {Biviano}, {Branchini}, {Carbone},
  {Cardone}, {Clements}, {Colombi}, {Conselice}, {Cresci}, {Deacon}, {Dunlop},
  {Fedeli}, {Fontanot}, {Franzetti}, {Giocoli}, {Garcia-Bellido}, {Gow},
  {Heavens}, {Hewett}, {Heymans}, {Holland}, {Huang}, {Ilbert}, {Joachimi},
  {Jennins}, {Kerins}, {Kiessling}, {Kirk}, {Kotak}, {Krause}, {Lahav}, {van
  Leeuwen}, {Lesgourgues}, {Lombardi}, {Magliocchetti}, {Maguire}, {Majerotto},
  {Maoli}, {Marulli}, {Maurogordato}, {McCracken}, {McLure}, {Melchiorri},
  {Merson}, {Moresco}, {Nonino}, {Norberg}, {Peacock}, {Pello}, {Penny},
  {Pettorino}, {Di Porto}, {Pozzetti}, {Quercellini}, {Radovich}, {Rassat},
  {Roche}, {Ronayette}, {Rossetti}, {Sartoris}, {Schneider}, {Semboloni},
  {Serjeant}, {Simpson}, {Skordis}, {Smadja}, {Smartt}, {Spano}, {Spiro},
  {Sullivan}, {Tilquin}, {Trotta}, {Verde}, {Wang}, {Williger}, {Zhao},
  {Zoubian}, \& {Zucca}}]{Euclid_redbook}
{Laureijs}, R., {Amiaux}, J., {Arduini}, S., {et~al.} 2011, arXiv e-prints,
  arXiv:1110.3193

\bibitem[{{Le Brun} {et~al.}(2015){Le Brun}, {McCarthy}, \&
  {Melin}}]{Le-Brun_2015}
{Le Brun}, A. M.~C., {McCarthy}, I.~G., \& {Melin}, J.-B. 2015, \mnras, 451,
  3868

\bibitem[{{Le Brun} {et~al.}(2014){Le Brun}, {McCarthy}, {Schaye}, \&
  {Ponman}}]{Le-Brun_2014}
{Le Brun}, A. M.~C., {McCarthy}, I.~G., {Schaye}, J., \& {Ponman}, T.~J. 2014,
  \mnras, 441, 1270

\bibitem[{{Lee} {et~al.}(2022){Lee}, {Coulton}, {Thiele}, \& {Ho}}]{Lee_2022}
{Lee}, B. K.~K., {Coulton}, W.~R., {Thiele}, L., \& {Ho}, S. 2022, \mnras, 517,
  420

\bibitem[{{Ma} {et~al.}(2015){Ma}, {Van Waerbeke}, {Hinshaw}, {Hojjati},
  {Scott}, \& {Zuntz}}]{Ma_2015}
{Ma}, Y.-Z., {Van Waerbeke}, L., {Hinshaw}, G., {et~al.} 2015, \jcap, 2015, 046

\bibitem[{{Maniyar} {et~al.}(2021){Maniyar}, {B{\'e}thermin}, \&
  {Lagache}}]{Maniyar_2021}
{Maniyar}, A., {B{\'e}thermin}, M., \& {Lagache}, G. 2021, \aap, 645, A40

\bibitem[{{McCarthy} {et~al.}(2018){McCarthy}, {Bird}, {Schaye},
  {Harnois-Deraps}, {Font}, \& {van Waerbeke}}]{McCarthy_2018}
{McCarthy}, I.~G., {Bird}, S., {Schaye}, J., {et~al.} 2018, \mnras, 476, 2999

\bibitem[{{McCarthy} {et~al.}(2014){McCarthy}, {Le Brun}, {Schaye}, \&
  {Holder}}]{McCarthy_2014}
{McCarthy}, I.~G., {Le Brun}, A.~M.~C., {Schaye}, J., \& {Holder}, G.~P. 2014,
  \mnras, 440, 3645

\bibitem[{{McCarthy} {et~al.}(2023){McCarthy}, {Salcido}, {Schaye}, {Kwan},
  {Elbers}, {Kugel}, {Schaller}, {Helly}, {Braspenning}, {Frenk}, {van Daalen},
  {Vandenbroucke}, {Conley}, {Font}, \& {Upadhye}}]{McCarthy_2023}
{McCarthy}, I.~G., {Salcido}, J., {Schaye}, J., {et~al.} 2023, \mnras, 526,
  5494

\bibitem[{{McCarthy} {et~al.}(2017){McCarthy}, {Schaye}, {Bird}, \& {Le
  Brun}}]{McCarthy_2017-bahamas}
{McCarthy}, I.~G., {Schaye}, J., {Bird}, S., \& {Le Brun}, A. M.~C. 2017,
  \mnras, 465, 2936

\bibitem[{{Mead} {et~al.}(2020){Mead}, {Tr{\"o}ster}, {Heymans}, {Van
  Waerbeke}, \& {McCarthy}}]{Mead_2020}
{Mead}, A.~J., {Tr{\"o}ster}, T., {Heymans}, C., {Van Waerbeke}, L., \&
  {McCarthy}, I.~G. 2020, \aap, 641, A130

\bibitem[{{Menanteau} {et~al.}(2010){Menanteau}, {Gonz{\'a}lez}, {Juin},
  {Marriage}, {Reese}, {Acquaviva}, {Aguirre}, {Appel}, {Baker}, {Barrientos},
  {Battistelli}, {Bond}, {Das}, {Deshpande}, {Devlin}, {Dicker}, {Dunkley},
  {D{\"u}nner}, {Essinger-Hileman}, {Fowler}, {Hajian}, {Halpern},
  {Hasselfield}, {Hern{\'a}ndez-Monteagudo}, {Hilton}, {Hincks}, {Hlozek},
  {Huffenberger}, {Hughes}, {Infante}, {Irwin}, {Klein}, {Kosowsky}, {Lin},
  {Marsden}, {Moodley}, {Niemack}, {Nolta}, {Page}, {Parker}, {Partridge},
  {Sehgal}, {Sievers}, {Spergel}, {Staggs}, {Swetz}, {Switzer}, {Thornton},
  {Trac}, {Warne}, \& {Wollack}}]{ACT_2010}
{Menanteau}, F., {Gonz{\'a}lez}, J., {Juin}, J.-B., {et~al.} 2010, \apj, 723,
  1523

\bibitem[{{Mo} \& {White}(1996)}]{Mo_1996}
{Mo}, H.~J. \& {White}, S.~D.~M. 1996, \mnras, 282, 347

\bibitem[{{Moser} {et~al.}(2022){Moser}, {Battaglia}, {Nagai}, {Lau}, {Machado
  Poletti Valle}, {Villaescusa-Navarro}, {Amodeo}, {Angl{\'e}s-Alc{\'a}zar},
  {Bryan}, {Dave}, {Hernquist}, \& {Vogelsberger}}]{Moser_2022}
{Moser}, E., {Battaglia}, N., {Nagai}, D., {et~al.} 2022, \apj, 933, 133

\bibitem[{{Osato} {et~al.}(2020){Osato}, {Shirasaki}, {Miyatake}, {Nagai},
  {Yoshida}, {Oguri}, \& {Takahashi}}]{Osato_2020}
{Osato}, K., {Shirasaki}, M., {Miyatake}, H., {et~al.} 2020, \mnras, 492, 4780

\bibitem[{{Osato} \& {Takada}(2021)}]{Osato_2021}
{Osato}, K. \& {Takada}, M. 2021, \prd, 103, 063501

\bibitem[{{Pandey} {et~al.}(2023){Pandey}, {Lehman}, {Baxter}, {Ni},
  {Angl{\'e}s-Alc{\'a}zar}, {Genel}, {Villaescusa-Navarro}, {Delgado}, \& {di
  Matteo}}]{Pandey_2023}
{Pandey}, S., {Lehman}, K., {Baxter}, E.~J., {et~al.} 2023, \mnras, 525, 1779

\bibitem[{{Perotto} {et~al.}(2023){Perotto}, {Adam}, {Ade}, {Ajeddig},
  {Andr{\'e}}, {Artis}, {Aussel}, {Barrena}, {Bartalucci}, {Beelen},
  {Beno{\^\i}t}, {Berta}, {Bing}, {Bourrion}, {Calvo}, {Catalano}, {De Petris},
  {D{\'e}sert}, {Doyle}, {Driessen}, {Ejlali}, {Ferragamo}, {Gomez}, {Goupy},
  {Hanser}, {Katsioli}, {K{\'e}ruzor{\'e}}, {Kramer}, {Ladjelate}, {Lagache},
  {Leclercq}, {Lestrade}, {Mac{\'\i}as-P{\'e}rez}, {Madden}, {Maury},
  {Mauskopf}, {Mayet}, {Monfardini}, {Moyer-Anin}, {Mu{\~n}oz-Echeverr{\'\i}a},
  {Paliwal}, {Pisano}, {Pointecouteau}, {Ponthieu}, {Pratt}, {Rev{\'e}ret},
  {Rigby}, {Ritacco}, {Romero}, {Roussel}, {Ruppin}, {Schuster}, {Sievers},
  {Tucker}, \& {Yepes}}]{Perotto_2023}
{Perotto}, L., {Adam}, R., {Ade}, P., {et~al.} 2023, arXiv e-prints,
  arXiv:2310.04553

\bibitem[{{Plagge} {et~al.}(2010){Plagge}, {Benson}, {Ade}, {Aird}, {Bleem},
  {Carlstrom}, {Chang}, {Cho}, {Crawford}, {Crites}, {de Haan}, {Dobbs},
  {George}, {Hall}, {Halverson}, {Holder}, {Holzapfel}, {Hrubes}, {Joy},
  {Keisler}, {Knox}, {Lee}, {Leitch}, {Lueker}, {Marrone}, {McMahon}, {Mehl},
  {Meyer}, {Mohr}, {Montroy}, {Padin}, {Pryke}, {Reichardt}, {Ruhl},
  {Schaffer}, {Shaw}, {Shirokoff}, {Spieler}, {Stalder}, {Staniszewski},
  {Stark}, {Vanderlinde}, {Vieira}, {Williamson}, \& {Zahn}}]{Plagge_2010}
{Plagge}, T., {Benson}, B.~A., {Ade}, P.~A.~R., {et~al.} 2010, \apj, 716, 1118

\bibitem[{{Planck Collaboration} {et~al.}(2014{\natexlab{a}}){Planck
  Collaboration}, {Ade}, {Aghanim}, {Armitage-Caplan}, {Arnaud}, {Ashdown},
  {Atrio-Barandela}, {Aumont}, {Aussel}, {Baccigalupi}, {Banday}, {Barreiro},
  {Barrena}, {Bartelmann}, {Bartlett}, {Battaner}, {Benabed}, {Beno{\^\i}t},
  {Benoit-L{\'e}vy}, {Bernard}, {Bersanelli}, {Bielewicz}, {Bikmaev}, {Bobin},
  {Bock}, {B{\"o}hringer}, {Bonaldi}, {Bond}, {Borrill}, {Bouchet}, {Bridges},
  {Bucher}, {Burenin}, {Burigana}, {Butler}, {Cardoso}, {Carvalho}, {Catalano},
  {Challinor}, {Chamballu}, {Chary}, {Chen}, {Chiang}, {Chiang}, {Chon},
  {Christensen}, {Churazov}, {Church}, {Clements}, {Colombi}, {Colombo},
  {Comis}, {Couchot}, {Coulais}, {Crill}, {Curto}, {Cuttaia}, {Da Silva},
  {Dahle}, {Danese}, {Davies}, {Davis}, {de Bernardis}, {de Rosa}, {de Zotti},
  {Delabrouille}, {Delouis}, {D{\'e}mocl{\`e}s}, {D{\'e}sert}, {Dickinson},
  {Diego}, {Dolag}, {Dole}, {Donzelli}, {Dor{\'e}}, {Douspis}, {Dupac},
  {Efstathiou}, {Eisenhardt}, {En{\ss}lin}, {Eriksen}, {Feroz}, {Finelli},
  {Flores-Cacho}, {Forni}, {Frailis}, {Franceschi}, {Fromenteau}, {Galeotta},
  {Ganga}, {G{\'e}nova-Santos}, {Giard}, {Giardino}, {Gilfanov},
  {Giraud-H{\'e}raud}, {Gonz{\'a}lez-Nuevo}, {G{\'o}rski}, {Grainge},
  {Gratton}, {Gregorio}, {Groeneboom}, {Gruppuso}, {Hansen}, {Hanson},
  {Harrison}, {Hempel}, {Henrot-Versill{\'e}}, {Hern{\'a}ndez-Monteagudo},
  {Herranz}, {Hildebrandt}, {Hivon}, {Hobson}, {Holmes}, {Hornstrup}, {Hovest},
  {Huffenberger}, {Hurier}, {Hurley-Walker}, {Jaffe}, {Jaffe}, {Jones},
  {Juvela}, {Keih{\"a}nen}, {Keskitalo}, {Khamitov}, {Kisner}, {Kneissl},
  {Knoche}, {Knox}, {Kunz}, {Kurki-Suonio}, {Lagache}, {L{\"a}hteenm{\"a}ki},
  {Lamarre}, {Lasenby}, {Laureijs}, {Lawrence}, {Leahy}, {Leonardi},
  {Le{\'o}n-Tavares}, {Lesgourgues}, {Li}, {Liddle}, {Liguori}, {Lilje},
  {Linden-V{\o}rnle}, {L{\'o}pez-Caniego}, {Lubin}, {Mac{\'\i}as-P{\'e}rez},
  {MacTavish}, {Maffei}, {Maino}, {Mandolesi}, {Maris}, {Marshall}, {Martin},
  {Mart{\'\i}nez-Gonz{\'a}lez}, {Masi}, {Massardi}, {Matarrese}, {Matthai},
  {Mazzotta}, {Mei}, {Meinhold}, {Melchiorri}, {Melin}, {Mendes}, {Mennella},
  {Migliaccio}, {Mikkelsen}, {Mitra}, {Miville-Desch{\^e}nes}, {Moneti},
  {Montier}, {Morgante}, {Mortlock}, {Munshi}, {Murphy}, {Naselsky}, {Nati},
  {Natoli}, {Nesvadba}, {Netterfield}, {N{\o}rgaard-Nielsen}, {Noviello},
  {Novikov}, {Novikov}, {O'Dwyer}, {Olamaie}, {Osborne}, {Oxborrow}, {Paci},
  {Pagano}, {Pajot}, {Paoletti}, {Pasian}, {Patanchon}, {Pearson}, {Perdereau},
  {Perotto}, {Perrott}, {Perrotta}, {Piacentini}, {Piat}, {Pierpaoli},
  {Pietrobon}, {Plaszczynski}, {Pointecouteau}, {Polenta}, {Ponthieu}, {Popa},
  {Poutanen}, {Pratt}, {Pr{\'e}zeau}, {Prunet}, {Puget}, {Rachen}, {Reach},
  {Rebolo}, {Reinecke}, {Remazeilles}, {Renault}, {Ricciardi}, {Riller},
  {Ristorcelli}, {Rocha}, {Rosset}, {Roudier}, {Rowan-Robinson},
  {Rubi{\~n}o-Mart{\'\i}n}, {Rumsey}, {Rusholme}, {Sandri}, {Santos},
  {Saunders}, {Savini}, {Schammel}, {Scott}, {Seiffert}, {Shellard},
  {Shimwell}, {Spencer}, {Stanford}, {Starck}, {Stolyarov}, {Stompor},
  {Sudiwala}, {Sunyaev}, {Sureau}, {Sutton}, {Suur-Uski}, {Sygnet}, {Tauber},
  {Tavagnacco}, {Terenzi}, {Toffolatti}, {Tomasi}, {Tristram}, {Tucci},
  {Tuovinen}, {T{\"u}rler}, {Umana}, {Valenziano}, {Valiviita}, {Van Tent},
  {Vibert}, {Vielva}, {Villa}, {Vittorio}, {Wade}, {Wandelt}, {White}, {White},
  {Yvon}, {Zacchei}, \& {Zonca}}]{Planck_2013_SZsource}
{Planck Collaboration}, {Ade}, P.~A.~R., {Aghanim}, N., {et~al.}
  2014{\natexlab{a}}, \aap, 571, A29

\bibitem[{{Planck Collaboration} {et~al.}(2014{\natexlab{b}}){Planck
  Collaboration}, {Ade}, {Aghanim}, {Armitage-Caplan}, {Arnaud}, {Ashdown},
  {Atrio-Barandela}, {Aumont}, {Baccigalupi}, {Banday}, {Barreiro}, {Bartlett},
  {Battaner}, {Benabed}, {Beno{\^\i}t}, {Benoit-L{\'e}vy}, {Bernard},
  {Bersanelli}, {Bielewicz}, {Bobin}, {Bock}, {Bonaldi}, {Bond}, {Borrill},
  {Bouchet}, {Bridges}, {Bucher}, {Burigana}, {Butler}, {Calabrese},
  {Cappellini}, {Cardoso}, {Catalano}, {Challinor}, {Chamballu}, {Chary},
  {Chen}, {Chiang}, {Chiang}, {Christensen}, {Church}, {Clements}, {Colombi},
  {Colombo}, {Couchot}, {Coulais}, {Crill}, {Curto}, {Cuttaia}, {Danese},
  {Davies}, {Davis}, {de Bernardis}, {de Rosa}, {de Zotti}, {Delabrouille},
  {Delouis}, {D{\'e}sert}, {Dickinson}, {Diego}, {Dolag}, {Dole}, {Donzelli},
  {Dor{\'e}}, {Douspis}, {Dunkley}, {Dupac}, {Efstathiou}, {Elsner},
  {En{\ss}lin}, {Eriksen}, {Finelli}, {Forni}, {Frailis}, {Fraisse},
  {Franceschi}, {Gaier}, {Galeotta}, {Galli}, {Ganga}, {Giard}, {Giardino},
  {Giraud-H{\'e}raud}, {Gjerl{\o}w}, {Gonz{\'a}lez-Nuevo}, {G{\'o}rski},
  {Gratton}, {Gregorio}, {Gruppuso}, {Gudmundsson}, {Haissinski}, {Hamann},
  {Hansen}, {Hanson}, {Harrison}, {Henrot-Versill{\'e}},
  {Hern{\'a}ndez-Monteagudo}, {Herranz}, {Hildebrandt}, {Hivon}, {Hobson},
  {Holmes}, {Hornstrup}, {Hou}, {Hovest}, {Huffenberger}, {Jaffe}, {Jaffe},
  {Jewell}, {Jones}, {Juvela}, {Keih{\"a}nen}, {Keskitalo}, {Kisner},
  {Kneissl}, {Knoche}, {Knox}, {Kunz}, {Kurki-Suonio}, {Lagache},
  {L{\"a}hteenm{\"a}ki}, {Lamarre}, {Lasenby}, {Lattanzi}, {Laureijs},
  {Lawrence}, {Leach}, {Leahy}, {Leonardi}, {Le{\'o}n-Tavares}, {Lesgourgues},
  {Lewis}, {Liguori}, {Lilje}, {Linden-V{\o}rnle}, {L{\'o}pez-Caniego},
  {Lubin}, {Mac{\'\i}as-P{\'e}rez}, {Maffei}, {Maino}, {Mandolesi}, {Maris},
  {Marshall}, {Martin}, {Mart{\'\i}nez-Gonz{\'a}lez}, {Masi}, {Massardi},
  {Matarrese}, {Matthai}, {Mazzotta}, {Meinhold}, {Melchiorri}, {Melin},
  {Mendes}, {Menegoni}, {Mennella}, {Migliaccio}, {Millea}, {Mitra},
  {Miville-Desch{\^e}nes}, {Moneti}, {Montier}, {Morgante}, {Mortlock}, {Moss},
  {Munshi}, {Murphy}, {Naselsky}, {Nati}, {Natoli}, {Netterfield},
  {N{\o}rgaard-Nielsen}, {Noviello}, {Novikov}, {Novikov}, {O'Dwyer},
  {Osborne}, {Oxborrow}, {Paci}, {Pagano}, {Pajot}, {Paladini}, {Paoletti},
  {Partridge}, {Pasian}, {Patanchon}, {Pearson}, {Pearson}, {Peiris},
  {Perdereau}, {Perotto}, {Perrotta}, {Pettorino}, {Piacentini}, {Piat},
  {Pierpaoli}, {Pietrobon}, {Plaszczynski}, {Platania}, {Pointecouteau},
  {Polenta}, {Ponthieu}, {Popa}, {Poutanen}, {Pratt}, {Pr{\'e}zeau}, {Prunet},
  {Puget}, {Rachen}, {Reach}, {Rebolo}, {Reinecke}, {Remazeilles}, {Renault},
  {Ricciardi}, {Riller}, {Ristorcelli}, {Rocha}, {Rosset}, {Roudier},
  {Rowan-Robinson}, {Rubi{\~n}o-Mart{\'\i}n}, {Rusholme}, {Sandri}, {Santos},
  {Savelainen}, {Savini}, {Scott}, {Seiffert}, {Shellard}, {Spencer}, {Starck},
  {Stolyarov}, {Stompor}, {Sudiwala}, {Sunyaev}, {Sureau}, {Sutton},
  {Suur-Uski}, {Sygnet}, {Tauber}, {Tavagnacco}, {Terenzi}, {Toffolatti},
  {Tomasi}, {Tristram}, {Tucci}, {Tuovinen}, {T{\"u}rler}, {Umana},
  {Valenziano}, {Valiviita}, {Van Tent}, {Vielva}, {Villa}, {Vittorio}, {Wade},
  {Wandelt}, {Wehus}, {White}, {White}, {Wilkinson}, {Yvon}, {Zacchei}, \&
  {Zonca}}]{Planck_2013-cosmo}
{Planck Collaboration}, {Ade}, P.~A.~R., {Aghanim}, N., {et~al.}
  2014{\natexlab{b}}, \aap, 571, A16

\bibitem[{{Planck Collaboration} {et~al.}(2016){Planck Collaboration},
  {Aghanim}, {Arnaud}, {Ashdown}, {Aumont}, {Baccigalupi}, {Banday},
  {Barreiro}, {Bartlett}, {Bartolo}, {Battaner}, {Battye}, {Benabed},
  {Beno{\^\i}t}, {Benoit-L{\'e}vy}, {Bernard}, {Bersanelli}, {Bielewicz},
  {Bock}, {Bonaldi}, {Bonavera}, {Bond}, {Borrill}, {Bouchet}, {Burigana},
  {Butler}, {Calabrese}, {Cardoso}, {Catalano}, {Challinor}, {Chiang},
  {Christensen}, {Churazov}, {Clements}, {Colombo}, {Combet}, {Comis},
  {Coulais}, {Crill}, {Curto}, {Cuttaia}, {Danese}, {Davies}, {Davis}, {de
  Bernardis}, {de Rosa}, {de Zotti}, {Delabrouille}, {D{\'e}sert}, {Dickinson},
  {Diego}, {Dolag}, {Dole}, {Donzelli}, {Dor{\'e}}, {Douspis}, {Ducout},
  {Dupac}, {Efstathiou}, {Elsner}, {En{\ss}lin}, {Eriksen}, {Fergusson},
  {Finelli}, {Forni}, {Frailis}, {Fraisse}, {Franceschi}, {Frejsel},
  {Galeotta}, {Galli}, {Ganga}, {G{\'e}nova-Santos}, {Giard},
  {Gonz{\'a}lez-Nuevo}, {G{\'o}rski}, {Gregorio}, {Gruppuso}, {Gudmundsson},
  {Hansen}, {Harrison}, {Henrot-Versill{\'e}}, {Hern{\'a}ndez-Monteagudo},
  {Herranz}, {Hildebrandt}, {Hivon}, {Holmes}, {Hornstrup}, {Huffenberger},
  {Hurier}, {Jaffe}, {Jones}, {Juvela}, {Keih{\"a}nen}, {Keskitalo}, {Kneissl},
  {Knoche}, {Kunz}, {Kurki-Suonio}, {Lacasa}, {Lagache}, {L{\"a}hteenm{\"a}ki},
  {Lamarre}, {Lasenby}, {Lattanzi}, {Leonardi}, {Lesgourgues}, {Levrier},
  {Liguori}, {Lilje}, {Linden-V{\o}rnle}, {L{\'o}pez-Caniego},
  {Mac{\'\i}as-P{\'e}rez}, {Maffei}, {Maggio}, {Maino}, {Mandolesi},
  {Mangilli}, {Maris}, {Martin}, {Mart{\'\i}nez-Gonz{\'a}lez}, {Masi},
  {Matarrese}, {Melchiorri}, {Melin}, {Migliaccio}, {Miville-Desch{\^e}nes},
  {Moneti}, {Montier}, {Morgante}, {Mortlock}, {Munshi}, {Murphy}, {Naselsky},
  {Nati}, {Natoli}, {Noviello}, {Novikov}, {Novikov}, {Paci}, {Pagano},
  {Pajot}, {Paoletti}, {Pasian}, {Patanchon}, {Perdereau}, {Perotto},
  {Pettorino}, {Piacentini}, {Piat}, {Pierpaoli}, {Pietrobon}, {Plaszczynski},
  {Pointecouteau}, {Polenta}, {Ponthieu}, {Pratt}, {Prunet}, {Puget}, {Rachen},
  {Reinecke}, {Remazeilles}, {Renault}, {Renzi}, {Ristorcelli}, {Rocha},
  {Rossetti}, {Roudier}, {Rubi{\~n}o-Mart{\'\i}n}, {Rusholme}, {Sandri},
  {Santos}, {Sauv{\'e}}, {Savelainen}, {Savini}, {Scott}, {Spencer},
  {Stolyarov}, {Stompor}, {Sunyaev}, {Sutton}, {Suur-Uski}, {Sygnet}, {Tauber},
  {Terenzi}, {Toffolatti}, {Tomasi}, {Tramonte}, {Tristram}, {Tucci},
  {Tuovinen}, {Valenziano}, {Valiviita}, {Van Tent}, {Vielva}, {Villa}, {Wade},
  {Wandelt}, {Wehus}, {Yvon}, {Zacchei}, \& {Zonca}}]{Planck_2015}
{Planck Collaboration}, {Aghanim}, N., {Arnaud}, M., {et~al.} 2016, \aap, 594,
  A22

\bibitem[{{Refregier} {et~al.}(2000){Refregier}, {Komatsu}, {Spergel}, \&
  {Pen}}]{Refregier_2000}
{Refregier}, A., {Komatsu}, E., {Spergel}, D.~N., \& {Pen}, U.-L. 2000, \prd,
  61, 123001

\bibitem[{{Refregier} \& {Teyssier}(2002)}]{Refregier_2002}
{Refregier}, A. \& {Teyssier}, R. 2002, \prd, 66, 043002

\bibitem[{{Sadeh} {et~al.}(2007){Sadeh}, {Rephaeli}, \& {Silk}}]{Sadeh_2007}
{Sadeh}, S., {Rephaeli}, Y., \& {Silk}, J. 2007, \mnras, 380, 637

\bibitem[{{Schaye} {et~al.}(2010){Schaye}, {Dalla Vecchia}, {Booth}, {Wiersma},
  {Theuns}, {Haas}, {Bertone}, {Duffy}, {McCarthy}, \& {van de
  Voort}}]{Schaye_2010}
{Schaye}, J., {Dalla Vecchia}, C., {Booth}, C.~M., {et~al.} 2010, \mnras, 402,
  1536

\bibitem[{{Seljak} {et~al.}(2001){Seljak}, {Burwell}, \& {Pen}}]{Seljak_2001}
{Seljak}, U., {Burwell}, J., \& {Pen}, U.-L. 2001, \prd, 63, 063001

\bibitem[{{Sheth} {et~al.}(2001){Sheth}, {Mo}, \& {Tormen}}]{Sheth_2001}
{Sheth}, R.~K., {Mo}, H.~J., \& {Tormen}, G. 2001, \mnras, 323, 1

\bibitem[{{Sheth} \& {Tormen}(1999)}]{Sheth_1999}
{Sheth}, R.~K. \& {Tormen}, G. 1999, \mnras, 308, 119

\bibitem[{{Sorini} {et~al.}(2024){Sorini}, {Bose}, {Pakmor}, {Hernquist},
  {Springel}, {Hadzhiyska}, {Hern{\'a}ndez-Aguayo}, \& {Kannan}}]{Sorini_2024}
{Sorini}, D., {Bose}, S., {Pakmor}, R., {et~al.} 2024, arXiv e-prints,
  arXiv:2409.01758

\bibitem[{{Spacek} {et~al.}(2018){Spacek}, {Richardson}, {Scannapieco},
  {Devriendt}, {Dubois}, {Peirani}, \& {Pichon}}]{Spacek_2018}
{Spacek}, A., {Richardson}, M. L.~A., {Scannapieco}, E., {et~al.} 2018, \apj,
  865, 109

\bibitem[{{Springel} \& {Hernquist}(2003)}]{Springel_2003}
{Springel}, V. \& {Hernquist}, L. 2003, \mnras, 339, 289

\bibitem[{{Springel} {et~al.}(2001){Springel}, {White}, \&
  {Hernquist}}]{Springel_2001}
{Springel}, V., {White}, M., \& {Hernquist}, L. 2001, \apj, 549, 681

\bibitem[{{Springel} {et~al.}(2005){Springel}, {White}, {Jenkins}, {Frenk},
  {Yoshida}, {Gao}, {Navarro}, {Thacker}, {Croton}, {Helly}, {Peacock}, {Cole},
  {Thomas}, {Couchman}, {Evrard}, {Colberg}, \& {Pearce}}]{Springel2005}
{Springel}, V., {White}, S. D.~M., {Jenkins}, A., {et~al.} 2005, \nat, 435, 629

\bibitem[{{Sunyaev} \& {Zeldovich}(1970)}]{Sunyaev1970}
{Sunyaev}, R.~A. \& {Zeldovich}, Y.~B. 1970, \apss, 7, 3

\bibitem[{{Sutherland} \& {Dopita}(1993)}]{Sutherland93}
{Sutherland}, R.~S. \& {Dopita}, M.~A. 1993, \apjs, 88, 253

\bibitem[{{Teyssier}(2002)}]{Teyssier_2002}
{Teyssier}, R. 2002, \aap, 385, 337

\bibitem[{{Tornatore} {et~al.}(2003){Tornatore}, {Borgani}, {Springel},
  {Matteucci}, {Menci}, \& {Murante}}]{Tornatore_2003}
{Tornatore}, L., {Borgani}, S., {Springel}, V., {et~al.} 2003, \mnras, 342,
  1025

\bibitem[{{Tornatore} {et~al.}(2006){Tornatore}, {Maier}, \&
  {Pattavina}}]{Tornatore_2006}
{Tornatore}, M., {Maier}, G., \& {Pattavina}, A. 2006, Journal of Optical
  Networking, 5, 858

\bibitem[{{Tr{\"o}ster} {et~al.}(2022){Tr{\"o}ster}, {Mead}, {Heymans}, {Yan},
  {Alonso}, {Asgari}, {Bilicki}, {Dvornik}, {Hildebrandt}, {Joachimi},
  {Kannawadi}, {Kuijken}, {Schneider}, {Shan}, {van Waerbeke}, \&
  {Wright}}]{Troster_2022}
{Tr{\"o}ster}, T., {Mead}, A.~J., {Heymans}, C., {et~al.} 2022, \aap, 660, A27

\bibitem[{{Van Waerbeke} {et~al.}(2014){Van Waerbeke}, {Hinshaw}, \&
  {Murray}}]{Van-Waerbeke_2014}
{Van Waerbeke}, L., {Hinshaw}, G., \& {Murray}, N. 2014, \prd, 89, 023508

\bibitem[{{Villaescusa-Navarro}(2018)}]{Pylians}
{Villaescusa-Navarro}, F. 2018, {Pylians: Python libraries for the analysis of
  numerical simulations}, Astrophysics Source Code Library, record
  ascl:1811.008

\bibitem[{{Vogelsberger} {et~al.}(2014){Vogelsberger}, {Genel}, {Springel},
  {Torrey}, {Sijacki}, {Xu}, {Snyder}, {Nelson}, \&
  {Hernquist}}]{Vogelsberger_2014}
{Vogelsberger}, M., {Genel}, S., {Springel}, V., {et~al.} 2014, \mnras, 444,
  1518

\bibitem[{{Waizmann} \& {Bartelmann}(2009)}]{Waizmann_2009}
{Waizmann}, J.~C. \& {Bartelmann}, M. 2009, \aap, 493, 859

\bibitem[{{Wiersma} {et~al.}(2009){Wiersma}, {Schaye}, {Theuns}, {Dalla
  Vecchia}, \& {Tornatore}}]{Wiersma_2009b}
{Wiersma}, R. P.~C., {Schaye}, J., {Theuns}, T., {Dalla Vecchia}, C., \&
  {Tornatore}, L. 2009, \mnras, 399, 574

\end{thebibliography}

\begin{appendix}
\section{Power spectrum computation} \label{app:power_spectrum}
\subsection{Power spectrum computation in the \texttt{Horizon} suite of simulations}
To obtain the different power spectra from the simulations, we followed different procedures for the \texttt{Horizon} and \texttt{Magneticum} suite of simulations. 
\subsubsection{Power spectrum computation in the \texttt{Horizon} suite of simulations}
To compute the power spectra in the \texttt{Horizon} suite of simulations, we first needed to project the component of interest onto a uniform three-dimensional grid. The matter component is the sum of the dark matter, stars, and gas. DM and stars were projected with a cloud-in-cell interpolation on the grid. 
Gas quantities were already on the regular Cartesian grid and we directly used the values of mass or pressure from the corresponding level of refinement. The simulations provided the total gas pressure, which we can easily modify to obtain the electron pressure, assuming local thermodynamic equilibrium between ions and electrons for a fully ionized gas, that is:
\begin{equation}
    \frac{P_{\rm e}}{P} = \frac{\mu}{\mu_{\rm e}} \simeq 0.492,
\end{equation}
where $P$ is the total gas pressure, $\mu$ ($\mu_{\rm e}$) is the mean molecular weight for gas (electron) particles.  We projected our quantity into a $512^3$ grid, allowing us to reach a Nyquist frequency ($k_\text{Ny} = \pi \times N_{\rm mesh}/L_{\rm box}$) of $k_\text{Ny} \sim 16\, h\,\text{Mpc}^{-1}$ for \texttt{Horizon-AGN} and \texttt{Horizon-noAGN} and up to $k_\text{Ny} \sim 1.8 \, h \,\text{Mpc}^{-1}$ for \texttt{Horizon-Large}. To obtain the angular power spectrum, it was beneficial to project \texttt{Horizon-Large} into a $1024^3$ grid to achieve $k_\text{Ny} \sim 3.6\, h\,\text{Mpc}^{-1}$.

Once the quantity is projected, we used the \texttt{Pylians} python package \citep{Pylians} to compute the 3D auto- and cross-power spectra deconvolved by the CIC mass-assignment scheme.

\subsection{Power spectrum computation in \texttt{Magneticum}}
For the \texttt{Magneticum} simulation, we assigned each gas particle (labelled by $i$) an electron pressure $P_{{\rm e}, i}$ according to the ideal gas law
\begin{align}
    P_{{\rm e}, i} &= \frac{N_{{\rm e},i} k_{\rm B} T_i}{V_{\mathrm{cell}}},\\
    N_{{\rm e},i} &= \frac{M_i}{m_{\rm p}\mu_{{\rm e}, i}},
\end{align}
where $T_i$ is the particle temperature, $V_{\mathrm{cell}}$ is the cell volume, $N_{{\rm e},i}$ is the number of free electrons, $M_i$ is the SPH particle mass, $\mu_{{\rm e},i}=2/(2-Y_i)$ is the mean mass per electron and $Y_i$ is the Helium fraction.

We used $\texttt{Pylians}$ to project the electron pressure onto a 1024$^3$ mesh based on the CIC assignment scheme and then measured the power spectrum. We thus achieve $k_\text{Ny} \sim 3.6\, h\,\text{Mpc}^{-1}$.

\end{appendix}
\end{document}